\documentclass[twocolumn]{aastex631}

\usepackage{graphicx}
\usepackage[caption=false]{subfig}

\begin{document}

\title{Periodic Variability of the Central Stars of Planetary Nebulae\\ 
Surveyed through the Zwicky Transient Facility}

\author[0009-0007-5623-2475]{Pinjian Chen}
\affiliation{CAS Key Laboratory of Optical Astronomy, National Astronomical Observatories, Chinese Academy of Sciences (NAOC), 20A Datun Road, Beijing 100101, P.~R.\ China}
\affiliation{School of Astronomy and Space Science, University of Chinese Academy of Sciences, Beijing, 100049, P.~R.\ China}

\author[0000-0003-1286-2743]{Xuan Fang}
\affiliation{CAS Key Laboratory of Optical Astronomy, National Astronomical Observatories, Chinese Academy of Sciences (NAOC), 20A Datun Road, Beijing 100101, P.~R.\ China}
\affiliation{School of Astronomy and Space Science, University of Chinese Academy of Sciences, Beijing, 100049, P.~R.\ China}
\affiliation{Xinjiang Astronomical Observatory, Chinese Academy of Sciences, 150 Science 1-Street, Urumqi, Xinjiang, 830011, P.~R.\ China}
\affiliation{Laboratory for Space Research, Faculty of Science, The University of Hong Kong, Pokfulam Road, Hong Kong, P.~R.\ China}

\author[0000-0001-7084-0484]{Xiaodian Chen}
\affiliation{CAS Key Laboratory of Optical Astronomy, National Astronomical Observatories, Chinese Academy of Sciences (NAOC), 20A Datun Road, Beijing 100101, P.~R.\ China}
\affiliation{School of Astronomy and Space Science, University of Chinese Academy of Sciences, Beijing, 100049, P.~R.\ China}
\affiliation{Institute for Frontiers in Astronomy and Astrophysics, Beijing Normal University, Beijing 102206, P.~R.\ China}

\author[0000-0002-2874-2706]{Jifeng Liu}
\affiliation{CAS Key Laboratory of Optical Astronomy, National Astronomical Observatories, Chinese Academy of Sciences (NAOC), 20A Datun Road, Beijing 100101, P.~R.\ China}
\affiliation{School of Astronomy and Space Science, University of Chinese Academy of Sciences, Beijing, 100049, P.~R.\ China}
\affiliation{Institute for Frontiers in Astronomy and Astrophysics, Beijing Normal University, Beijing 102206, P.~R.\ China}
\affiliation{New Cornerstone Science Laboratory, National Astronomical Observatories, Chinese Academy of Sciences (NAOC), Beijing 100101, P.~R.\ China}

\correspondingauthor{Xuan Fang}
\email{fangx@nao.cas.cn}

\begin{abstract}
A consensus has been reached in recent years that binarity plays an important role in the formation and evolution of a significant fraction of planetary nebulae (PNe).  Utilizing the archived photometric data from the Zwicky Transient Facility survey, we conducted a comprehensive data mining in search for brightness variations in a large sample of Galactic PNe. This effort leads to identification of 39 PNe, whose central stars exhibit periodic variation in light curves. Among these objects, 20 are known binary central stars of PNe, while the remaining 19 are new discoveries. Additionally, we identified 14 PNe with central stars displaying anomalous variation in light curves, as well as eight variables based on the high-cadence photometric data.  Among the new discoveries of periodicity, we found compelling evidence of binary systems at the centres of two archetypal quadrupolar PNe. We also report on very peculiar brightness variation observed in the central core of the compact PN NGC\,6833. Several PNe in our sample deserve follow-up observations, both high-dispersion spectroscopy and high-precision photometry, to reveal the true nature of their central binarity or even multiplicity. 
\end{abstract}


\keywords{Planetary nebulae (1249) --- Planetary nebulae nuclei (1250) --- Binary stars (154) --- Close binary stars (254) --- Stellar evolution (1599)}

\section{Introduction} \label{sec:intro}

In the simple textbook version of the story, planetary nebulae (PNe) are descendants of the low- to intermediate-mass stars ($\sim$1--8\,$M_{\odot}$), which are among the most common inhabitants throughout the universe.  However, it has become clear in recent years that binarity plays an important role in the formation and evolution of a significant fraction of PNe \citep[e.g.][]{Miszalski2009,Jones2017,Jacoby2021}.  Formation of PNe through binary interaction has been proposed by numerical simulations of common envelope shaping of PNe \citep[e.g.][]{2018ApJ...860...19G,2020ApJ...893..150G}, and directly confirmed with the discovery of binary central stars of planetary nebulae (bCSPNe; hereafter in this paper, ``central stars of planetary nebulae'' is also written as CSPNe for short).  Several key and classical problems in astrophysics are also strongly linked to the binary evolution, including, but not limited to, the origin of PN central stars with H-deficient atmospheres \citep[e.g.][]{Miszalski2015,Jacoby2020}, the renowned ``abundance discrepancy'' problem in the deep spectroscopy of Galactic PNe \citep[e.g.][]{Corradi2015,Jones2016,Wesson2018}, and the extremely diverse, complex morphological strucutures \citep[i.e.\ elliptical, bipolar, multipolar, and irregular;][]{1998AJ....116.1357S,Sahai_2011,Parker2006} of PNe as revealed in high-resolution imaging \citep[e.g.][]{Hillwig2016}.  Though the details and extent of how binarity plays a role in PNe formation remain uncertain, it is no longer suitable to interpret this kind of astrophysical phenomenon solely with the scenario of single-star evolution. 

Among the most dramatic deviations from single-star evolution that binarity introduces is the common-envelope (CE) phase, during which the engulfment of the secondary results in ejection of the envelope (which later forms the nebular material of a PN), leaving a hot pre-white dwarf as the source of UV radiation to photoionize the nebula.  Therefore, the CE evolution is crucial for our understanding of the formation and evolution of PNe.  Conversely, given the relatively short lifetime ($\sim$10$^{4}$\,yr) of PNe, the system does not have enough time for subsequent binary evolution.  The relics of its past history should still be inscribed in the morphology and kinematics of PNe, making them ideal tracers for studying the CE evolution.  Moreover, the CE phase directly links PNe to other binary phenomena of particular interest, such as binary neutron stars \citep[e.g.][]{Ablimit2024} and Type~Ia supernovae \citep{Tsebrenko2015,2015Natur.519...63S,Court2024}.

\begin{figure*}[t]\centering 
    \includegraphics[width=1.5\columnwidth]{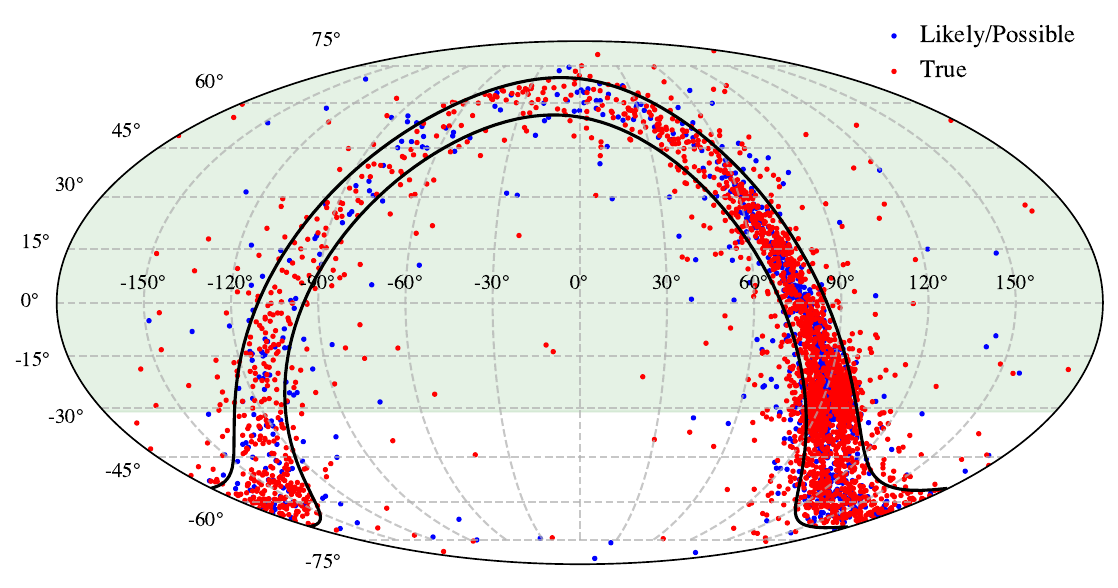}
     \caption{Spatial distribution of PNe and candidates in HASH.  Those labeled as ``True'' are marked as red dots, and the blue dots represent objects labeled as ``Likely'' or ``Possible''. The green shadow regions roughly define the sky coverage of ZTF ($\delta >-31^{\circ}$). The black lines correspond to $|b|=7^{\circ}$.
    \label{fig:pn distribution}}
\end{figure*}

Time-resolved photometry has proven to be an efficient and powerful method for identifying bCSPNe through photometric variations, in particular for systems with short orbital periods \citep{Boffin2019}.  Since the discovery of the first bCSPNe \citep[the central star of Abell\,63;][]{Bond1976}, concerted efforts have been made in recent decades to substantially increase the sample of confirmed binary central systems of PNe with the help of modern photometric surveys using the ground-based and space-borne facilities such as OGLE \citep{Miszalski2009}, \emph{Gaia} \citep{Chornay2021gaia,Chornay2022}, \emph{Kepler}/\textit{K2} \citep{DeMarco2015,Jacoby2020,Jacoby2021}, \emph{TESS} \citep{Aller2020,Aller2024} and ZTF \citep{Bhattacharjee2024}. Additionally, surveys of long-term radial velocity monitoring have also contributed a relatively small but important sample of long-period bCSPNe \citep{VanWinckel2014,Jones2017AA,Miszalski2018}, for which our understanding of the population is still limited.  To date, the total number of the confirmed bCSPNe exceeds one hundred\footnote{Based on the list compiled, as of late 2023, at \url{https://www.drdjones.net/bcspn/}, contributed by David Jones.}.  However, both observations and theory suggest that this number probably represents only the tip of an iceberg.  There are still valuable archived survey data, especially those from the ground-based observations, that have not been fully explored. 

In this paper, we report a catalog of Galactic PNe, whose central stars are periodic variables, through careful data mining of the recently released Zwicky Transient Facility (ZTF) photometric survey.  The new generation extremely wide-field camera with mosaicked CCDs enables capturing dynamical changes of the universe with unprecedented depth and precision.  With ZTF's deep and long-term photometric data covering almost the entire northern sky, it is significantly promising to survey the photometric variability of a comprehensive sample of CSPNe in the Milky Way.  We have collected the most complete sample of Galactic PNe.  The light curve of each object (when available) was visually examined, and analyzed for periodicities.  Section\,\ref{sec:ztf} describes the observational data of the ZTF survey utilized in the data-mining procedure, which described in Section\,\ref{sec: search procedure}.  In Section\,\ref{sec: results}, we present the results of variable central stars of PNe and discuss the general properties of the ZTF bCSPNe sample in Section\,\ref{sec: discussion}, along with discussion on the individual objects that we consider as peculiar.  The conclusion and future perspectives are given in Section \ref{sec: conclusion}.

\begin{figure*}[t]\centering
    \includegraphics[width=1.9\columnwidth]{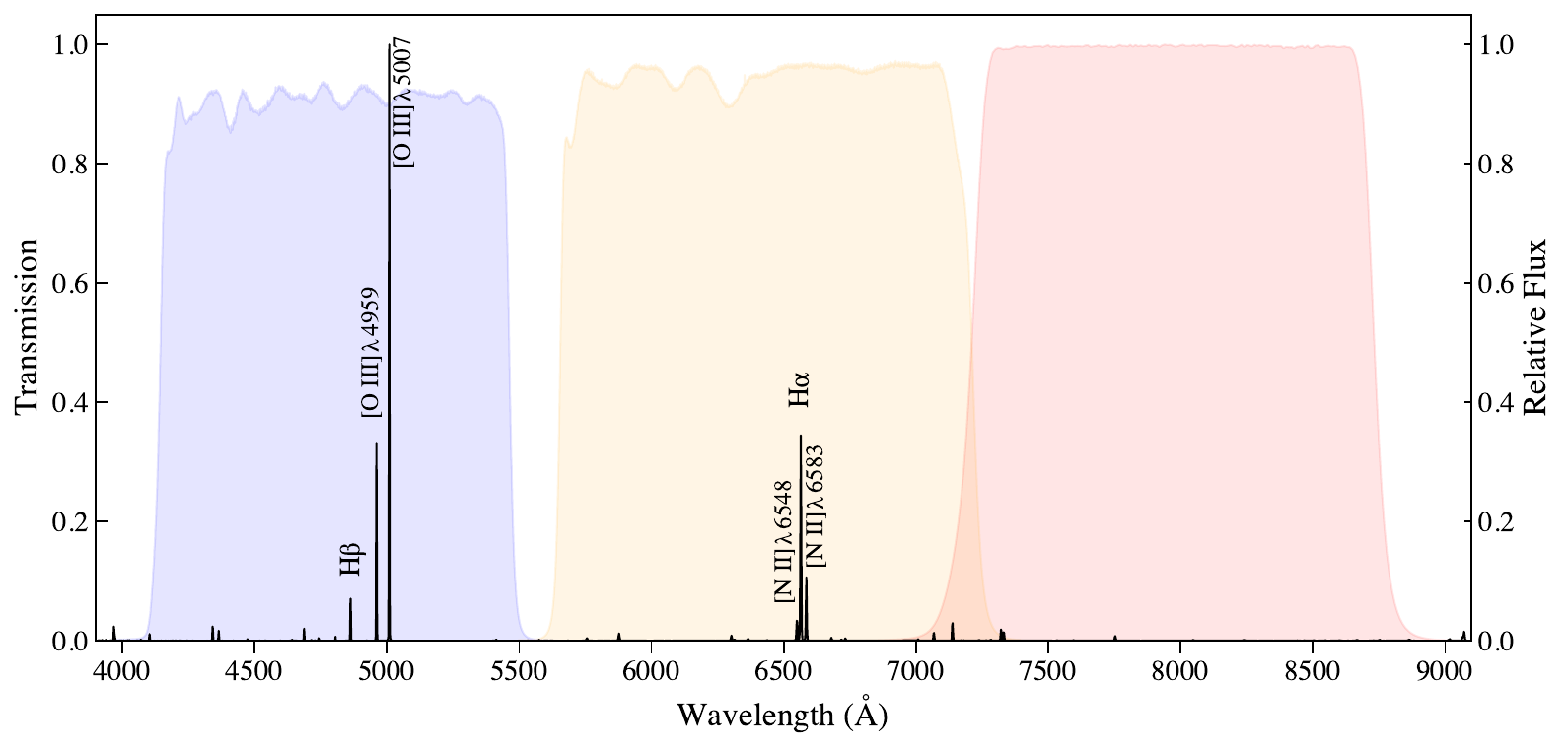}
     \caption{Filter transmissions for the ZTF $g$-, $r$-, and $i$-band filters (blue, orange, and red shadows), superposed on a low-resolution (R\,$\sim$\,1800) spectrum of PN K 3-69 from the Large Sky Area Multi-Object Fiber Spectroscopic Telescope \citep[LAMOST;][]{Cui2012, Zhao2012} database. The relative flux of the spectrum has been normalized to the level of [O~{\sc iii}] $\lambda$5007. The most prominent nebular emission lines are labeled.
    \label{fig: bandpass}}
\end{figure*}

\section{Sample and Data} \label{sec:ztf}

\subsection{The Galactic PNe Sample} \label{subsec: the galactic pn sample}

We collected the sample of Galactic PNe and candidates, along with their information of positions and angular diameters, based on the Hong Kong/AAO/Strasbourg H$\alpha$ (HASH) Planetary Nebula Database \citep{Parker2016,Bojicic_2016}.  In this catalog, there are 2,725 objects labeled as ``True'' (or T), and another 1,186 labeled as ``Likely'' (or L) or ``Possible (or P)''.  The majority of these objects are located in the Galactic disk; only those positioned north of $\delta$ = $-$31$^{\circ}$ can be possibly captured by ZTF (see Figure\,\ref{fig:pn distribution}).  It's worth mentioning that there are inevitably contentious objects, such as the PG\,1034$+$001 (Hewett~1), which was considered as a genuine PN by \citet{Aller2020} instead of being defined as ionized interstellar medium (ISM) in HASH. Such objects might be missed in our work.

\subsection{The ZTF Photometry} \label{subsec: ztf photometry} 

The ZTF is an optical time-domain survey utilizing the 48-inch Schmidt Telescope at Palomar Observatory \citep{Bellm2019,Masci2019}.  The equipped wide-field camera with a 600 megapixel CCD mosaic provides a 47\,$\rm deg^2$ field of view (FoV) and 8\,$\rm s$ readout time, making ZTF a powerful facility of detecting various photometric variables of scientific significance in the universe \citep{Graham2019}. Since the ZTF survey began in 2018 March 17, 40\% of the available observing time during Phase-I (2018 March -- 2020 September) and 50\% of the observing time during Phase-II (starting 2020 December) are allocated for public surveys.
In the public observations, there are two major surveys: a Northern Sky Survey and a Galactic Plane Survey.  About 27,500\,$\rm deg^2$ of the survey-fields in the visible northern sky are scanned with a three-day cadence (in Phase-I) and a two-day cadence (in Phase-II), and those fields close to the Galactic plane ($|b|\leq$\,7$^{\circ}$ and $\delta\,>$ $-$31$^{\circ}$) are observed nightly in Phase-I.  In both surveys, the $g$- and $r$-band photometry are obtained with a typical limiting magnitude of $g\sim$20.8\,mag and $r\sim$20.6\,mag, respectively. The standard ZTF exposure time is 30 seconds. In this work, we adopted the archival ZTF DR20, which provides access to the public survey data obtained from 2018 March to 2023 October, as well as data acquired under private survey time between 2018 March and 2022 June, such as the ZTF high-cadence Galactic plane survey \citep{Kupfer2021}.  The $i$-band photometry is available in part of the private surveys. 

A large catalog of periodic variables was reported in the pioneering effort of \citet{Chen2020} based on the ZTF data.  In this work, we narrow our focus on the photometric variability of the central stars of PNe.  Compared to other modern photometric surveys, ZTF has unique advantages in the search for bCSPNe.  First, the extremely large sky coverage and long-term photometry of ZTF produce observations of a number of Galactic PNe and candidates that are sufficient for statistical studies.  Second, multi-band photometry is conducted in the ZTF survey, which enables us to verify the accuracy of the period determined in different bands.  It also provides crucial constraints on the stellar temperatures of the binary system, which cannot be derived from a monochromatic light curve.  Additionally, the 1.0\arcsec\ pixel$^{-1}$ scale and the median angular resolution of 2\arcsec\ full-width at half-maximum (FWHM) effectively reduce the possible contamination from the background/foreground sources. 

 The irregular sampling pattern and the long-baseline observation ensure that ZTF is capable of detecting short timescale periodicity, even down to levels of several minutes \citep[e.g.][]{Burdge2019, Burdge2020a, Burdge2020b, Caiazzo2021}. However, ground-based observations are inevitably limited by complex weather and atmospheric conditions, which directly reduce the measurement precision. Consequently, compared to space missions such as \emph{TESS} and \emph{Kepler}/\emph{K2}, the sensitivity of ZTF photometry is still limited, hampering its detection of the photometric variability with small amplitudes. Another major concern in the photometry of CSPNe is possible contamination of nebular emission (in particular for angularly compact PNe), which cannot be overstated.  In bright PNe, the luminous and irregular nebular emission may drown the periodic signals from the central star and, at times, introduces spurious variability.  This issue highlights the importance of choosing appropriate filters in photometry.  As depicted in Figure\,\ref{fig: bandpass}, the ZTF $g$-band photometry suffers contamination from strong H$\beta$ and [O~{\sc iii}] nebular emission, while the $r$ band can be contaminated by the prominent H$\alpha$ and [N~{\sc ii}] $\lambda\lambda$6548,6584 nebular emission lines.  The optimal choice is the $i$ band where the nebular emission is minimal (Figure\,\ref{fig: bandpass}).  Unfortunately, the ZTF $i$-band observations have fewer epochs than in the other two bands, as this band is only used for private surveys.

\section{Searching Procedure} 
\label{sec: search procedure}

\subsection{Data Preparation} 
\label{subsec: data preparation}

Among the 3,911 PNe and candidates in the HASH database, 2,504 are situated within the sky coverage of ZTF. All the ZTF photometric data (in the $g$, $r$, and $i$ bands, where available) within a 5\arcsec\ radius of the geometric centres of these objects (as provided by HASH) were retrieved from the NASA/IPAC Infrared Science Archive (IRSA\footnote{\url{https://irsa.ipac.caltech.edu/Missions/ztf.html}}) using the application program interface (API).  Here the 5\arcsec\ radius from the PN centre is a reasonable estimate given that the central stars of PNe might be displaced from the geometric centre of nebulae.  For example, the central star of Abell\,46 is situated approximately 4\farcs1 from the geometric centre of the PN.  However, this possible displacement of the central star from the geometric centre of a PN introduces a challenge in distinguishing photometric variation of the CSPN from emission of the background and/or foreground variable stars, a matter to be addressed in the subsequent discussion. 

Since the ZTF observing system is fully robotic, poor-quality or unusable photometric measurements caused by various reasons such as scattered moonlight, cloudy weather etc., might pollute the observed light curves of targets. Therefore we specifically selected data with the flags \texttt{catflags}=0 and \texttt{magerr}$>$0 to ensure quality of photometric data, as recommended in the ZTF Science Data System (ZSDS) Explanatory Supplement\footnote{\url{https://irsa.ipac.caltech.edu/data/ZTF/docs/ztf_explanatory_supplement.pdf}} and the public data release notes\footnote{\url{https://irsa.ipac.caltech.edu/data/ZTF/docs/releases/dr20/ztf_release_notes_dr20.pdf}}.
 
Each object observed by ZTF in a specific filter is characterised by an object ID (OID). However, a unique source may be associated with multiple OIDs if it was observed in different ZTF fields, or with different CCDs and CCD quadrants. This introduces complexity when aggregating all light curves for a unique source. Furthermore, as the majority of PNe are located within the Galactic plane, it is common for nearby objects (e.g.\ field stars) to be included within a 5\arcsec\ radius. To address this issue, we decided to treat each light curve with different OIDs independently, even if they may originate from the same source. Additionally, we excluded light curves with fewer than 20 epochs to ensure the reliability of the data.  We designated the OID, which is associated with the maximum number of observation epochs, as the ``primary OID".  As a rough estimate of how ZTF observations cover our sample, we calculated the median number of observation epochs linked to the ``primary OID" after applying quality cuts. The resulting median values across the $g$, $r$, and $i$ bands are 209, 463, and 52, respectively.

\subsection{Visual Examination} \label{subsec: visual examination}

The ZTF light curves in the $g$, $r$ and $i$ bands were first plotted and visually examined.  This step aims to identify objects with anomalous behaviors in light curves, such as large-amplitude, long-term, semi-regular or irregular variations, and ``mysterious outbursts''.  Various mechanisms may drive these behaviors, including binary interaction, contamination by PNe mimics (e.g.\ symbiotic star, hereafter SySt), stellar winds, mass-loss and accretion events; even conditions of observations might cause notable phenomenons in light curves. 

Most of the PNe and candidates in our sample are located in the Galactic plane, and thus some of them were included in the ZTF high-cadence Galactic plane survey \citep{Kupfer2021}. This project aims to detect variable stars with periods shorter than a few hours, such as ultra-compact binaries, by conducting continuous $r$-band observations from 1.5 to 6\,hr with a cadence of 40\,s in the sky regions of low Galactic latitudes. The identification of fast, aperiodic variabilities can benefit from such dense sampling rates. When available, we extracted the high-cadence photometric data separately, and plotted and carefully examined each light curve.

\subsection{Search for Periodic Variables} \label{subsec: search for periodic variables}

We employed the \texttt{Astropy} \citep{astropy1,astropy2} implementation of the Lomb-Scargle method \citep{Lomb1976, Scargle1982, VanderPlas2018} to identify periodic variables.  As stated in Section\,\ref{subsec: data preparation}, each light curve associated with a unique OID was treated separately as a distinct source.  Periodograms were generated individually for each object and filter over a period range of 0.025\,d to 1500\,d, utilizing standard normalization.  To ensure that each peak in the periodogram is sampled with sufficient precision, we set the peak oversampling factor to 150. For each period, the false-alarm probability (FAP) was calculated and recorded using the method proposed by \citet{Baluev2008} to assess the likelihood of a period being discovered due to coincidental alignment. We collected suspected variable candidates by adopting a criterion of $\rm{log(FAP)}<-$3. Further visual examination of these individuals helped us to exclude aliased periods caused by the sampling cadence of ZTF, which strongly contaminate our results with periods of 1~day and its multiples.  Ultimately, only those objects exhibiting strong signatures of periodicity were retained.

Although several other PNe displayed potential signs of periodicity, their phase-folded light curves are noisy and irregular, likely affected by factors such as nebular emission and other systematic issues.  These objects were excluded from the final sample selection. For example, we detected a period of approximately 6.5~d in the $g$ band for the central star of LoTr\,1, which aligns roughly with the rotation period identified by \citet{Tyndall2013}.  However, the irregular shape of its light curve, likely due to sparse sampling, make us exclude this object in the subsequent analysis. Some of these objects will be further discussed in Section \ref{subsec: comparison with other studies}.

\subsection{Identification of PN Central Stars} \label{subsec: central star identification}

The aim of this step is to confirm that the periodic variability originates from the PN central star rather than from field stars.  We first downloaded the ZTF deep reference images of each periodic variable identified in the previous step, which typically combined 100 single science images. We then computed the reference position for each object in three passbands (when available) using the median values of right ascension and declination from ZTF observations, and adopted the passband with the highest number of epochs. Now with the angular diameter information and the reference position, we required each source to be projected within the coverage area of the nebula and near its geometric centre.  Our examination indicated that, in most cases where angular diameter information was available, the angular separation was negligible compared to the angular size of the nebula.  However, we observed a few cases where the absolute angular separation values are non-negligible (i.e., larger than 1\arcsec).

To further investigate the origin of the periodic variability, we consulted existing catalogs of CSPNe. Recently, \citet[hereafter CW21]{Chornay2021} and \citet[hereafter GS21]{Gonzalez-Santamaria2021} presented two catalogs of CSPNe from \emph{Gaia} EDR3. By considering both the $G_{\mathrm{BP}}-G_{\mathrm{RP}}$ colour and the angular separation $r$ between the object and the PN geometric centre, they identified more than 2,000 CSPNe with varying confidence levels. There are slight differences between the two catalogs. We performed a cross-match between our results and the two catalogs with a radius of 1\arcsec. Among the periodic variables, 38 matched with CW21, while 37 matched with GS21. The difference is due to NGC\,6833, which successfully cross-matched with CW21 but is absent from GS21. All matched objects exhibit an angular separation smaller than 0.5\arcsec, confirming that the variability indeed originates from the central star. For the four periodic variables that failed to match the central star, we provide brief comments in the text below.

\begin{figure}[t]
\centering
\includegraphics[width=1.0\linewidth]{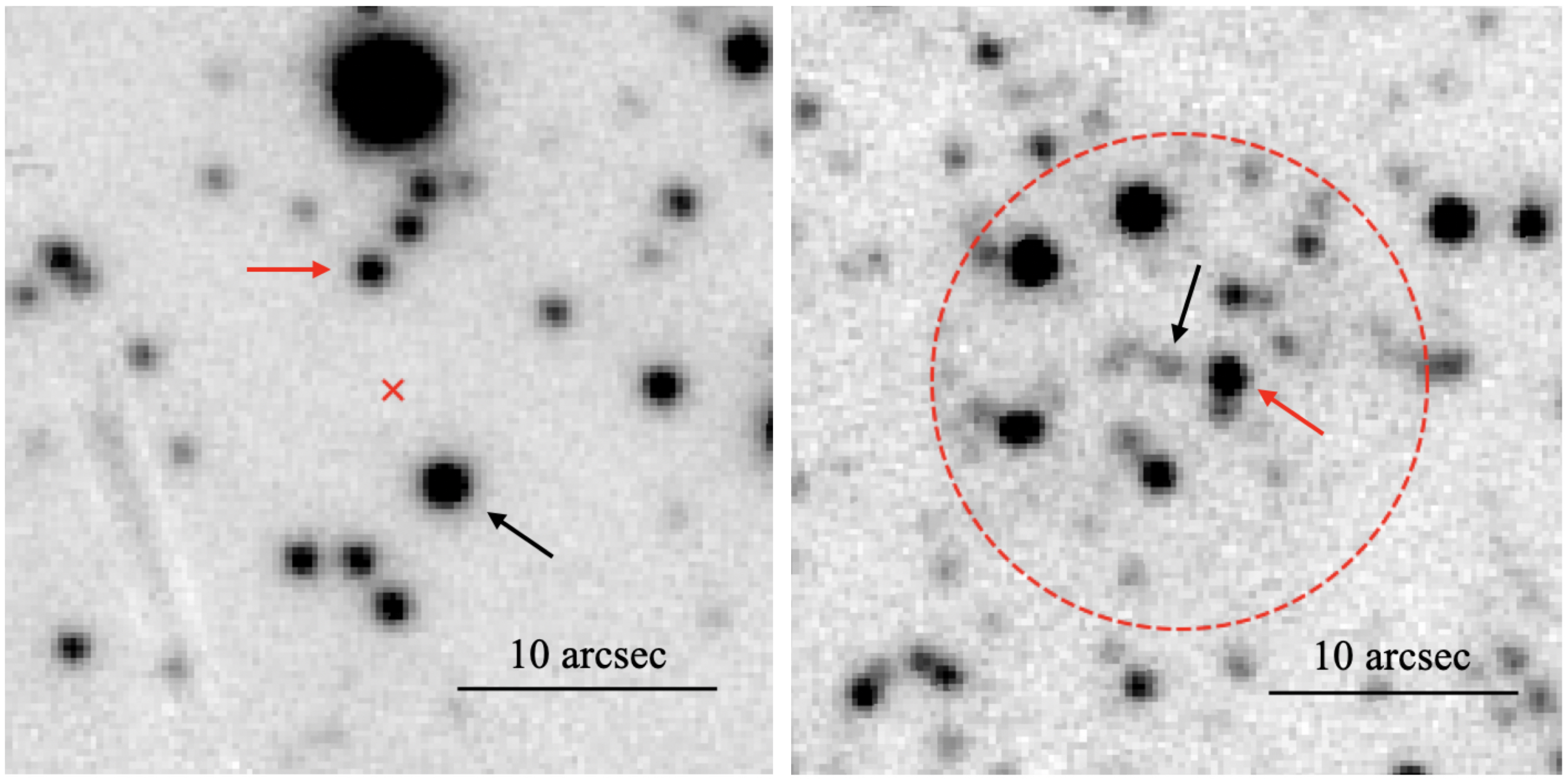} 
\caption{PanSTARRS $g$-band images of the PN candidate IPHASX\,J193636.3$+$123758 (left) and the genuine PN PHR\,J1738$-$2052 (right).  North is up and east to the left.  Periodic variables identified by ZTF are indicated by red arrows, and black arrows mark the positions of PN central stars identified by either CW21 or GS21.  The red cross in the left panel marks the geometric centre of IPHASX\,J193636.3$+$123758, while the red-dashed circle in the right panel represents the angular diameter of PHR\,J1738$-$2052.
    \label{fig: PS1}}
\end{figure}

\emph{IPHASX\,J193636.3+123758}: The variable identified by ZTF is located $\sim$4\farcs7 away from the centre of the nebula. However, the angular diameter of this PN candidate is 270\arcsec, making this displacement interpretable.  The central star of this PN candidate (``Likely" in HASH) is cataloged in GS21, but not in CW21. GS21 identified a bluer and brighter central star which is $\sim$4\arcsec\ away from the centre of the nebula (see Figure \ref{fig: PS1}).  Due to a lack of sufficient information, we keep this variable as the possible central star of this PN.

\emph{StDr\,7}:  This nebula is labeled as a ``Possible'' PN by HASH, and not included in either CW21 or GS21. With an angular diameter of 18\arcsec, we identified a periodic variable located $\sim$3\farcs1 away from the centre.  After image checking, we found no potential central stars near the nebular centre.  We keep this variable as the possible central star.

\emph{PPA\,J1805-2503}:  This object is labeled as a ``Likely" PN in HASH and is not included in either CW21 or GS21. We are confident that the periodic variability originates from this candidate PN due to its point-source nature, despite the lack of angular diameter information.  However, whether this object is a genuine PN remains uncertain. For this step, we retain the object for subsequent analysis. 

\emph{PHR\,J1738-2052}:  A genuine PN with a round morphology.  We found a short-period variable that is located $\sim$1\farcs7 away from the centre of the nebula.  However, both CW21 and GS21 identified another star as the central star, which is located $\sim$0\farcs9 away from the centre (see Figure \ref{fig: PS1}). The variable is also known as a Delta~Scuti Variable \citep{Soszynski2021}, reported as OGLE BLG-DSCT-12850 in SIMBAD.  The central star identified by CW21 and GS21 is bluer and fainter than the variable.  Therefore, we conclude that the variable is a field star coincidentally projected within the coverage area of PHR\,J1738-2052.  This object is no longer included in our discussion.

\subsection{Final Confirmation of Periods} \label{subsec: ultimate period confirmation}
Suboptimal observing conditions will lead to less good image quality and reduce the astrometric accuracy \citep{Masci2019}.  In the context of our study, these limitations may lead to increased nebular contamination, potentially hindering precise period determination. Generally, this becomes more evident at greater angular distances from the central star. Therefore, we recalculated the periods for all periodic variables, using all of the photometric data within the 1\arcsec \,radius around their positions, as provided in $Gaia$ DR3. However, detailed visual inspection indicated that stricter criteria may be required for specific cases, such as Kn\,26 (see Section \ref{subsec: periodic variables}). Additionally, this process combined observations associated with different OIDs for each object. For Fr\,2-21, the period of 0.9647~d in both the $g$ and $r$ bands disappeared after combining data from two OIDs, indicating that it is likely an alias introduced by ZTF's sampling cadence. Consequently, we excluded this object from our results.

\begin{table*}[p]
\setlength{\tabcolsep}{2.2pt}
\caption{\label{table: bcspne} ZTF Detections of Variable PNe}
\begin{center}
    \begin{tabular}{llccrccccc}
    \hline\hline
     Name$^a$ & PNG & RA & Dec  & Period & Passband & Type & HASH$^b$ & Morphology$^c$ & Class\\
     ~ & ~ & (J2000) & (J2000) & (d) & & & status & \\
     \hline 
        JaSt\,2-4* & 001.0+01.4 & 17:42:28.1 & -27:13:32.16 &  0.4661 & $g,r$ & Ell & T & E & A\\
        PPA\,J1805-2503 & 005.5-01.8 & 18:05:20.3 & -25:03:32.30 & 276.7289 & $r$ & ? & L & ~ & C\\
        CGMW\,4-1723 & 006.2-09.1 & 18:35:44.6 & -27:51:21.06 & 0.1298 & $r$ & Ell: & T & E & A\\ 
        PTB\,26* & 008.3+09.6 & 17:29:13.3 & -16:47:43.60 & 0.1522 & $g,r$ & Ell, Ecl & T & B & A\\ 
        Abell\,41* & 009.6+10.5 & 17:29:02.0 & -15:13:05.20  & 0.2265 & $g$ & Ell & T & B & A\\ 
        MA\,2 & 022.5+04.8 & 18:15:13.4 & -06:57:12.20 & 0.2596 & $i$ & Irr: & T & E & A\\ 
        M\,2-46 & 024.8-02.7 & 18:46:34.6 & -08:28:01.85 & 0.3192 & $g$ & Ell, Ecl & T & Q & A\\ 
        PHR\,J1847+0132 & 033.8+01.5 & 18:47:45.4 & +01:32:47.00 & 0.1267 & $r,i$ & Ell & T & B & A\\
        Pa\,13 & 041.4-09.6 & 19:41:21.0 & +03:07:17.90 & 0.3988 & $g,r$ & Ecl & T & E & A\\
        IPHASX\,J193636.3+123758 & 049.3-04.0 & 19:36:36.5 & +12:38:02.27 & 2.2150 & $r$ & ? & L & R & B\\ 
        Hen\,2-428* & 049.4+02.4 & 19:13:05.5 & +15:46:39.36 &  0.1757 & $g,r$ & Ell, Ecl & T & B & A\\ 
        IPHASX\,J194359.5+170901* & 054.2-03.4 & 19:43:59.5 & +17:09:01.08 &  1.1614 & $g,r,i$ & Irr & T & B & A\\ 
        Abell\,46* & 055.4+16.0 & 18:31:18.3 & +26:56:12.90 & 0.4717 & $g,r$ & Irr, Ecl & T & E & A\\ 
        NGC\,6905 & 061.4-09.5 & 20:22:22.9 & +20:06:16.81 &  651.9731 & $r$ & ? & T & B & C\\ 
        Pa\,164* & 061.5-02.6 & 19:57:23.3 & +23:52:48.36 & 1.1688 & $g,r,i$ & Irr, Ecl & L & & A\\ 
        Fr\,2-16 & 062.9-25.2 & 21:18:18.7 & +12:01:32.30 & 1.3198 & $g,r$ & Irr & P & & A\\ 
        StDr\,7 & 063.2+00.2 & 19:50:11.3 & +26:43:32.09 & 2.6927 & $r,i$ & Irr & P & & B\\ 
        ETHOS\,1* & 068.1+11.0 & 19:16:31.4 & +36:09:47.88 & 0.5351 & $g,r$ & Irr & T & B & A\\ 
        IRAS\,19581+3320 & 070.0+01.8a & 20:00:06.9 & +33:29:01.00 & 41.3705 & $g,r$ & ? & P & & C\\ 
        Pa\,27* & 075.0-07.2 & 20:48:58.4 & +32:18:14.80 & 7.3678 & $g$ & Rot & T & E & A\\ 
        NGC\,6833 & 082.5+11.3 & 19:49:46.6 & +48:57:40.10 & 957.8991 & $r,i$ & ? & T & B & C\\ 
        Kn\,26 & 084.6-07.9 & 21:23:09.4 & +38:58:13.08  & 0.0490 & $g,r$ & Irr: & T & Q & A\\ 
        Ou\,5* & 086.9-03.4 & 21:14:20.0 & +43:41:36.00  & 0.3642 & $g,r,i$ & Irr, Ecl & T & B & A\\ 
        KTC\,1* & 099.1+05.7 & 21:28:11.0 & +58:52:34.68 & 1.3291 & $g,r$ & Irr & T & E & A\\ 
        Cr\,1* & 100.3+02.8 & 21:49:11.7 & +57:27:19.70 & 0.3569 & $g,r,i$ & Irr & P & E & A\\
        K\,1-6* & 107.0+21.3 & 20:04:14.3 & +74:25:36.00 & 21.3051 & $g$ & Rot & T & E & A\\ 
        IsWe\,2 & 107.7+07.8 & 22:13:22.5 & +65:53:55.40 & 0.7913 & $g,r$ & Irr: & T & E & A\\
        WeBo\,1* & 135.6+01.0 & 02:40:14.4 & +61:09:16.70 & 4.6964 & $g,r$ & Rot & T & B & A\\
        TS\,1 & 135.9+55.9 & 11:53:24.7 & +59:39:56.88 & 0.1635 & $g,r,i$ & Ell & T & E & A\\
        HFG\,1* & 136.3+05.5 & 03:03:47.0 & +64:54:35.40 & 0.5816 & $g,r$ & Irr & T & E & A\\
        LTNF\,1* & 144.8+65.8 & 11:57:44.8 & +48:56:18.38  & 2.2912 & $g,r,i$ & Irr, Ecl & T & B & A\\ 
        WPS\,54* & 162.1+47.9 & 09:51:25.9 & +53:09:30.71 & 3.4532 & $g,r,i$ & Rot: & P & & B\\
        Kn\,51 & 164.8-09.8 & 04:25:26.9 & +35:06:07.80 & 0.2688 & $g,r$ & Irr: & T & I & B\\
        CoMaC\,2 & 168.4+01.6 & 05:21:57.0 & +39:31:02.50 & 4.5877 & $r$ & Irr & T & B & A\\
        Kn\,133 & 173.7-09.2 & 04:54:33.7 & +28:49:28.92 & 2.4182 & $g,r$ & Irr & L & & A\\
        HaWe\,8 & 192.5+07.2 & 06:40:09.7 & +21:25:01.70 & 0.9868 & $g,r,i$ & Irr & T & E & A\\
        WeDe\,1* & 197.4-06.4 & 05:59:24.9 & +10:41:40.40 & 0.5707 & $g,r$ & Rot: & L & E & B\\
        PM\,1-23* & 222.8-04.2 & 06:54:13.4 & -10:45:38.30 & 1.2628 & $g,r$ & Ell & T & B & A\\
        MPA\,J0705-1224 & 225.5-02.5 & 07:05:37.2 & -12:24:51.91 & 0.1844 & $g,r$ & Ell & L & E & A\\
        \hline
    \end{tabular}
    \end{center}
    \tablenotetext{a}{Previous known bCSPNe are marked with asterisks, see detailed references from the website contributed by David Jones.}
    \tablenotetext{b}{Based on the labels from HASH, as of 20th June 2024. T: True; L: Likely; P: Possible.}
    \tablenotetext{c}{Morphology flag from HASH. E: Elliptical/oval; R: Round; B: Bipolar; I: Irregular. Q: Quadrupolar \citep{Manchado1996, Guerrero2013}.}
\end{table*}

\section{Results} \label{sec: results} 

\subsection{Periodic Variables} \label{subsec: periodic variables}
Table \ref{table: bcspne} lists the 39 PNe, whose central stars exhibit evident periodic variation as identified through the ZTF photometry.  Among these, 20 objects were previously known bCSPNe (with different levels of confidence), which was marked with asterisks, and the remaining 19 are new discoveries. The table also presents basic information about the central stars and their host PNe, including names, coordinates, photometric periods, the ZTF passbands that we utilized for period determination, labels from HASH, and morphology information of the PNe (when available). Generally, the periodic variation can be recovered from more than one passband, except in a few cases due to an insufficient number of data points or other intrinsic physical reasons (e.g. nebular contamination). 

The ``Type'' column in Table\,\ref{table: bcspne} describes the related photometric modulations.  In addition to common eclipsing systems (Ecl), this includes irradiation effects (Irr) due to the differing projection of the irradiated hemisphere of a cool secondary, ellipsoidal modulations (Ell) resulting from tidal distortions in close-binary systems, and rotation effects (Rot) from the rotation of a cool companion. It should be noted that other mechanisms, such as the Doppler beaming effect, can also contribute to the observed periodic variability. However, its level (amplitude $\leq$0.1\%) is too low to be detailed in the context of ZTF-like ground-based observations \citep{Zucker2007, Shporer2010}. We adopted similar criteria to those used in previous studies for classification, considering shape, period, and amplitude (see details in individual cases). Generally, ellipsoidal modulation is characterized by a relatively short period and small amplitude, whereas irradiation effects can exhibit a broader range of these values. Furthermore, the amplitude of the irradiation effect is filter-dependent, reflecting the effective temperature of the cool companion.

For each periodic variable, the best period returned by the periodogram might actually be half of its true orbital period.  We thus doubled the best period identified in the periodogram and examined the result. If the light curve displayed two minima of different depths, we attributed the variability to ellipsoidal modulation; otherwise, if it was nearly sinusoidal, we tentatively classified the variability as due to irradiation. For known bCSPNe with high-quality data, we referenced results from the existing literature. We emphasize that the classification for those scattered and low-amplitude light curves are tentative due to data limitations. For example, future high-accuracy photometric or radial velocity monitoring might better distinguish ellipsoidal variables from irradiated variables, thus doubling the orbital period. In Table\,\ref{table: bcspne}, uncertain classification results are marked with ``:'', and a few unusual detections with unclear driving mechanisms are denoted with ``?''.

Are all of these objects exhibiting periodic variability genuine bCSPNe? Based on the available ZTF light curves, we categorized these periodic variables into three classes (see the last column in Table \ref{table: bcspne}). Specifically, Class A includes high-confidence bCSPNe, whose periodic photometric variations are believed to strongly linked to the central binarity. Class B represents objects with lower confidence, due to uncertainties either in the identification of the central star or in the classification of their light curves. Finally, Class C encompasses those objects whose variability origin remains unclear and may not be associated with central binarity.  In total, there are 30, 5, and 4 variables classified into Classes A, B, and C, respectively.

\begin{figure*}
\centering
\includegraphics[width=2.1\columnwidth]{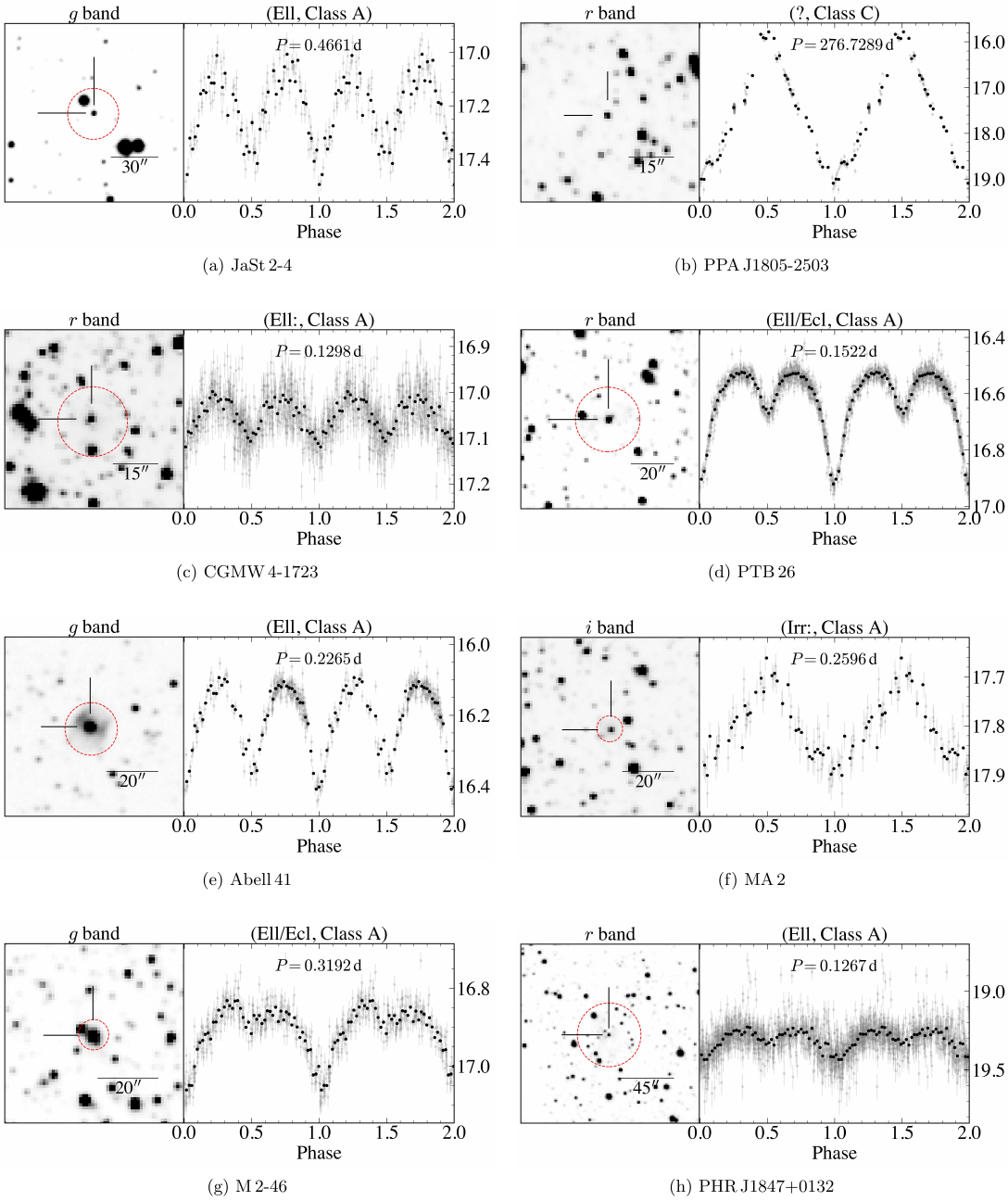}
\caption{Images and phase-folded ZTF light curves for (a) JaSt\,2-4, (b) PPA\,J1805-2503, (c) CGMW\,4-1723, (d) PTB\,26, (e) Abell\,41, (f) MA\,2, (g) M\,2-46, (h) PHR\,J1847+0132. The reference images are taken in the same passband as the corresponding light curves, as indicated in the titles. The angular diameter of each PN is shown by a red dashed circle, and the type of variability, classification, and period values are annotated (see text for details).}
  \label{fig: LCs1}
\end{figure*}

\begin{figure*}
\centering
\includegraphics[width=2.1\columnwidth]{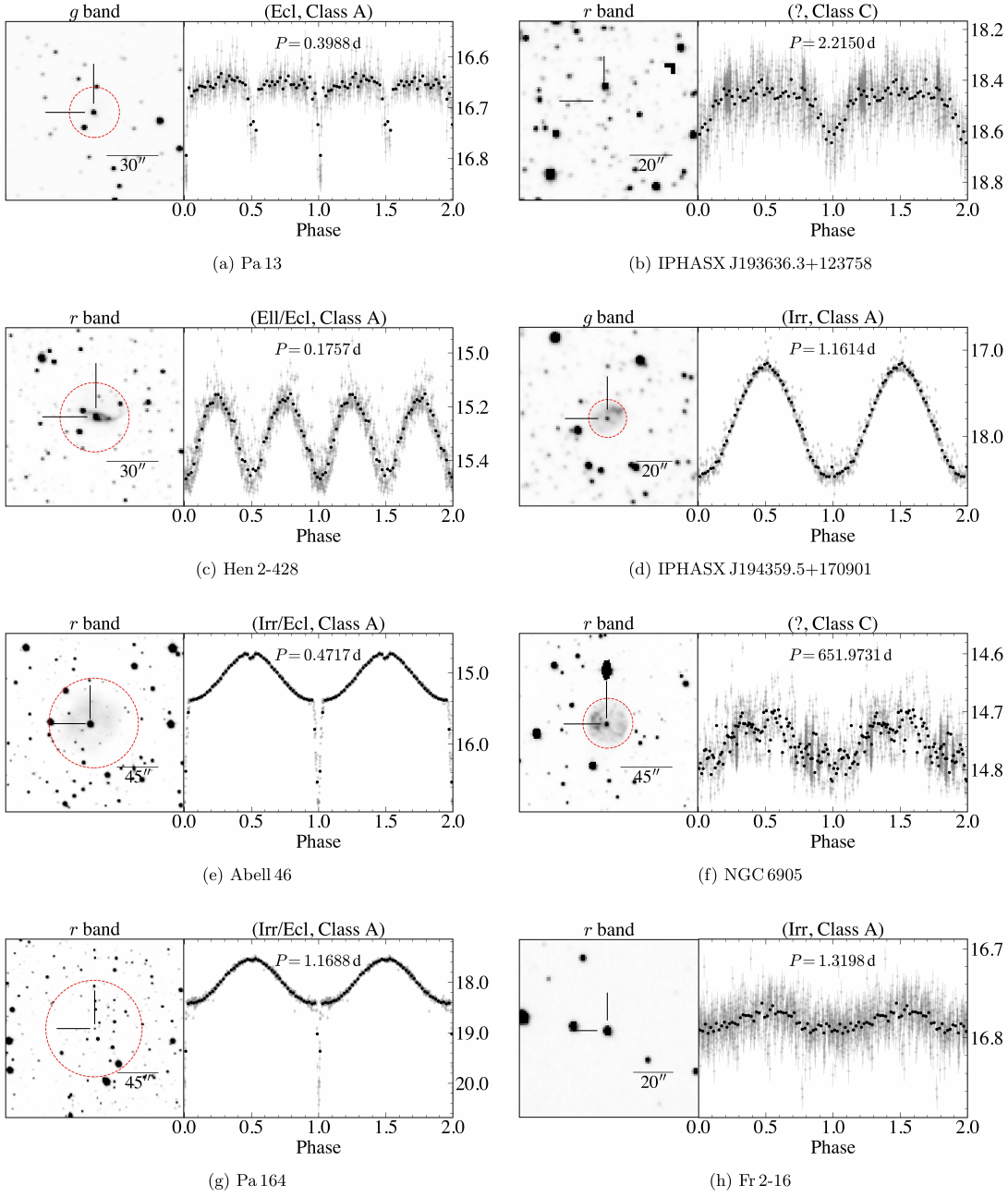}
\caption{Same as Figure\,\ref{fig: LCs1} but images and phase-folded ZTF light curves for (a) Pa\,13, (b) IPHASX\,J193636.3+123758, (c) Hen\,2-428, (d) IPHASX\,J194359.5+170901, (e) Abell\,46, (f) NGC\,6905, (g) Pa\,164, (h) Fr\,2-16.}
  \label{fig: LCs2}
\end{figure*}

\begin{figure*}
\centering
\includegraphics[width=2.1\columnwidth]{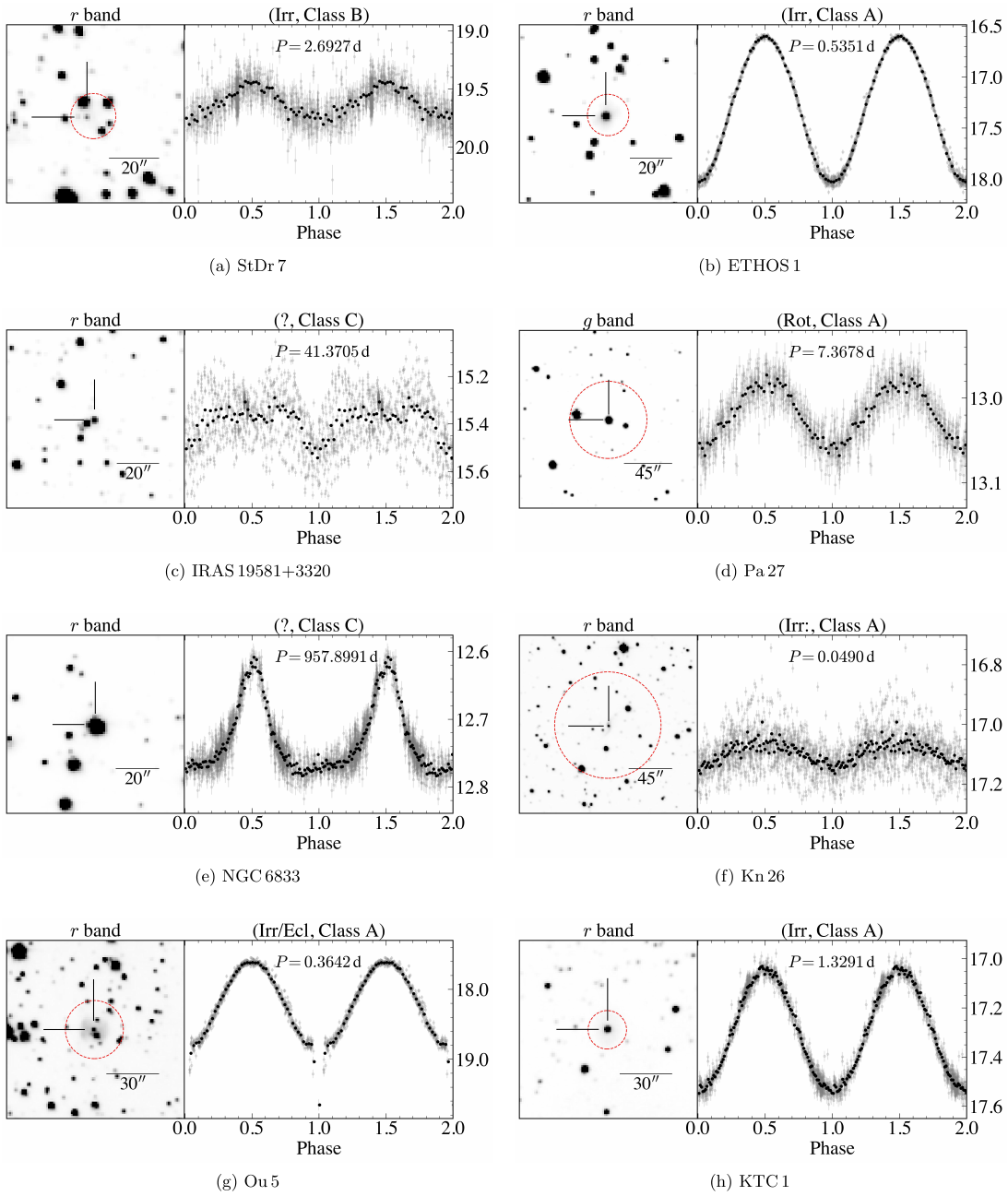} 
\caption{Same as Figure\,\ref{fig: LCs1} but images and phase-folded ZTF light curves for (a) StDr\,7, (b) ETHOS\,1, (c) IRAS\,19581+3320, (d) Pa\,27, (e) NGC\,6833, (f) Kn\,26, (g) Ou\,5, (h) KTC\,1.}
  \label{fig: LCs3}
\end{figure*}

\begin{figure*}
\centering
\includegraphics[width=2.1\columnwidth]{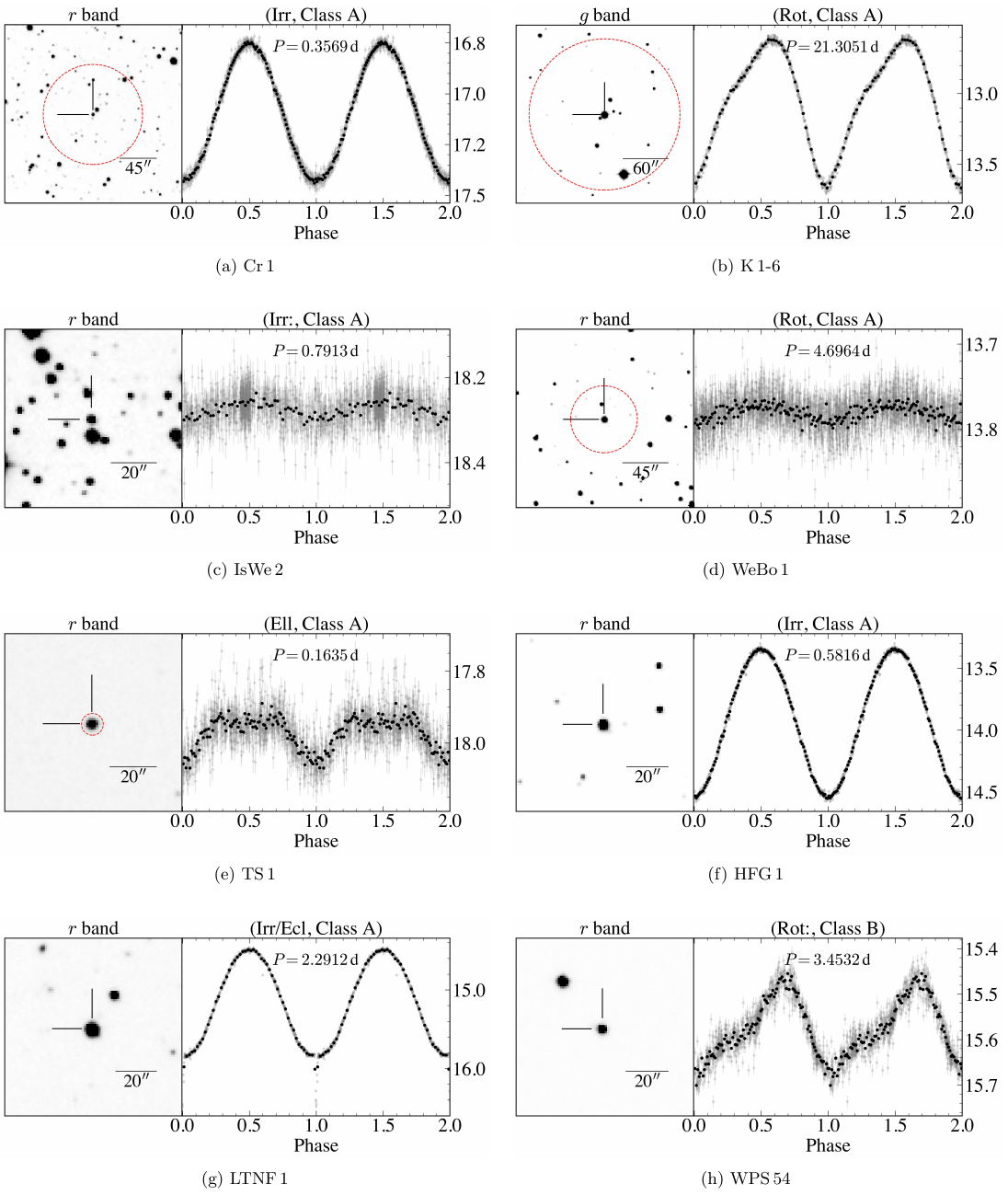}
\caption{Same as Figure\,\ref{fig: LCs1} but images and phase-folded ZTF light curves for (a) Cr\,1, (b) K\,1-6, (c) IsWe\,2, (d) WeBo\,1, (e) TS\,1, (f) HFG\,1, (g) LTNF\,1, (h) WPS\,54.}
  \label{fig: LCs4}
\end{figure*}

\begin{figure*}
\centering
\includegraphics[width=2.1\columnwidth]{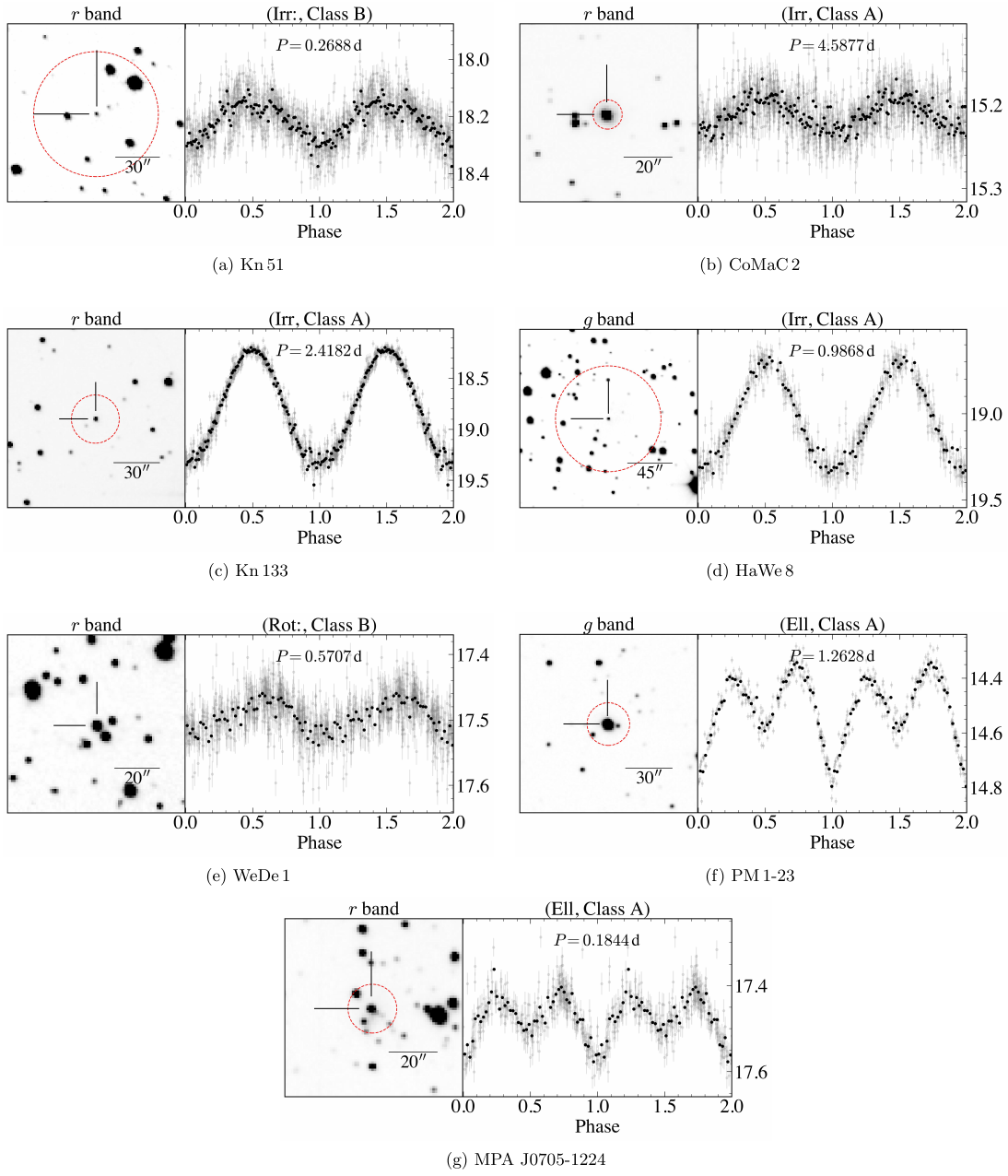}
\caption{Same as Figure\,\ref{fig: LCs1} but images and phase-folded ZTF light curves for (a) Kn\,51, (b) CoMaC\,2, (c) Kn\,133, (d) HaWe\,8, (e) WeDe\,1, (f) PM\,1-23, (g) MPA J0705-1224.}
  \label{fig: LCs5}
\end{figure*}

\begin{figure*}
\centering
\includegraphics[width=2.1\columnwidth]{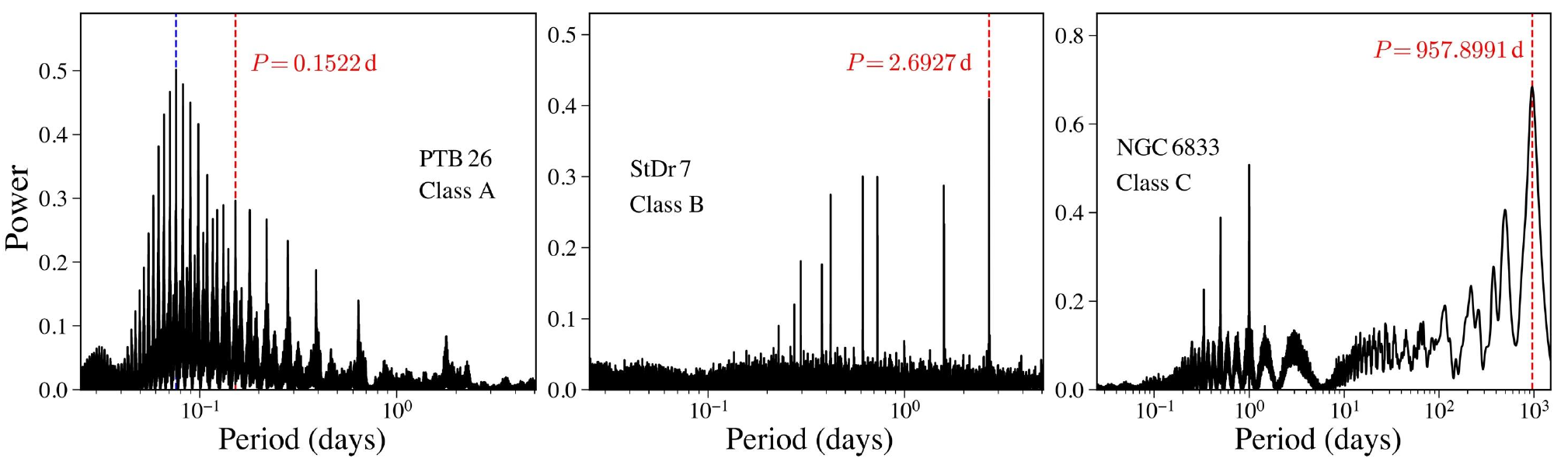}
\caption{Example periodograms for objects from Class A, B, and C (from left to right: PTB\,26, StDr\,7, and NGC\,6833), generated by the $r$-band light cuvres. The adopted periods are marked by red dashed lines with their values indicated. For PTB\,26, the highest peak in the periodogram (traced by the blue dashed line) corresponds to half the true orbital period.
 \label{fig: periodograms}}
\end{figure*}

In Figures\,\ref{fig: LCs1}--\ref{fig: LCs5}, we present the phase-folded light curves along with the corresponding ZTF deep reference images for all 39 variables, in the same order as they appear in Table\,\ref{table: bcspne}.  We adopted the ZTF passband that holds the most observations or displays the variability most clearly.  The folded light curves are averaged with a bin size of 0.01 or 0.02 in phase (based on the number of observations) to improve visibility.  The image size was chosen based on the angular diameter of the PN, which is depicted by a red dashed circle.  All images are adjusted with North to top and East to left.  The position of the variable is denoted by the intersection of two solid black lines. In some cases we do not show the sky coverage of the nebula either due to a lack of available angular diameter information or because the diameter is too large to be included in the image. In the special case of NGC\,6833, the nebular diameter is too small (0.6\arcsec) to be displayed. Additionally, we display three representative examples of periodograms in Figure\,\ref{fig: periodograms} for objects from Class A, B, and C.

\smallskip

Here we briefly discuss some of these periodic variables, with a particular focus on the newly discovered ones.

\emph{PPA\,J1805-2503}:  This object is classified as a long-period variable by \emph{Gaia} DR3 variability classification.  Although the number of $g$-band observations is insufficient, we identified a clear periodicity near 277~d in the $r$-band.  The long period, large amplitude (greater than 3 magnitudes), and approximate triangular shape are all unusual for typical bCSPNe.  We conclude that binarity might not be responsible for the observed variability, given the late-type star features present in its red spectrum (as provided by HASH).  Instead, we suspect that this ``Likely" PN is a symbiotic binary, considering the late-type star features evident in its red spectrum (provided by HASH).

\emph{CGMW\,4-1723}: The observation epochs in the $g$ band are insufficient, and the periodogram based on the 527 data points in the $r$ band produces a series of peaks introduced by the window function, with periods ranging from approximately 0.05 to 0.1~d. The highest peak corresponds to a period of 0.0649\,d. By doubling this period, we arrived at a light curve with two minima of slightly different depths, albeit with significant noise. We tentatively classify the variability as due to the ellipsoidal modulation with the data at hand, though the short orbital period and the limited number of observations hinder a definitive classification. Follow-up precise photometric or radial velocity observations are needed to refine the period estimate.

\emph{PTB\,26}: The label of PTB\,26 has recently been updated from ``Likely" to a ``True" PN in HASH. The photometric variation of its central star was reported by \cite{Chornay2021gaia} and \cite{Jacoby2021}, with the latter identifying a period of about 0.1522~d based on light curves from \emph{Kepler}/\emph{K2} and ATLAS. Due to the combined influence of the window function and the presence of harmonics, its periodogram exhibits a series of peaks from approximately 0.05 to 0.2~d (see Figure\,\ref{fig: periodograms}). We identified a consistent period based on the highest peak from both $g$- and $r$-band photometry. Intriguingly, the depths of the primary eclipse displayed in the ZTF $g$- and $r$-band light curves are deeper than those observed in the \emph{Kepler}/\emph{K2} light curve presented by \citet{Jacoby2021}. This may result from the difference in passbands, indicating a companion star with a relatively low temperature.

\emph{MA\,2}: Although there are nearly no $g$- and $r$-band observations of this object, we fortunately discovered a strong periodicity at 0.2596~d in the $i$ band. We tentatively classify the variability as due to the irradiation effect. Given the limited number of data points from ZTF, follow-up observations are needed to further improve the period estimate.

\begin{figure*}[t]
\centering
\includegraphics[width=1.5\columnwidth]{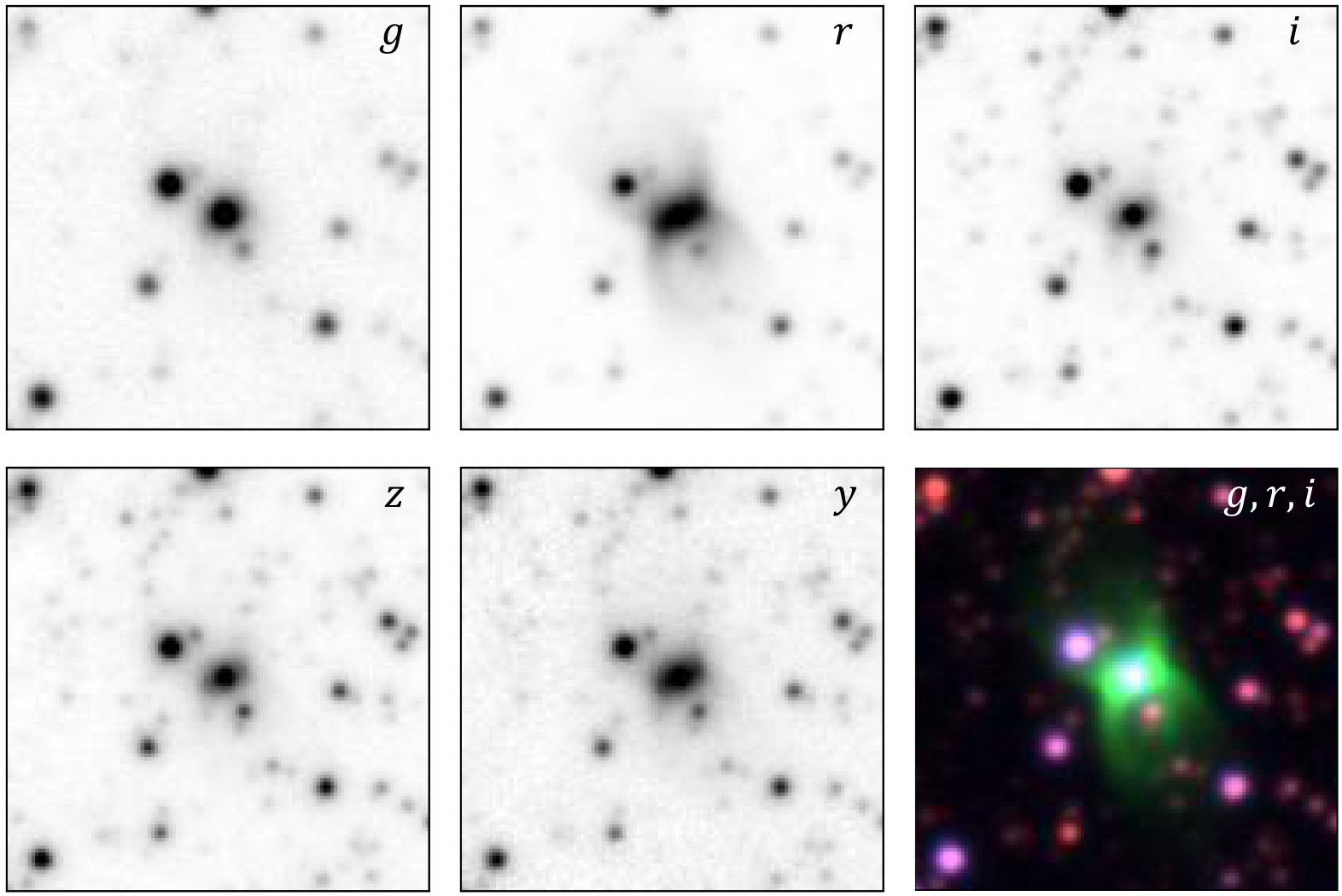}
\caption{PanSTARRS images of M\,2-46 in five bands ($g$, $r$, $i$, $z$, and $y$).  North is up and east to the left; image FoV is $\sim$30\arcsec$\times$30\arcsec.  M\,2-46 is an archetypal example showing how nebular contamination covers up the evidence of binarity when utilizing the data from modern photometric surveys.
 \label{fig: M 2-46}}
\end{figure*}

\emph{M\,2-46}:  This PN is among the five quadrupolar PNe reported by \citet{Manchado1996}.  We identified a period of 0.3192~d from the ZTF $g$-band photometry, but found nothing in the $r$ band despite a greater number of observations.  This can be attributed to the very low-excitation nature of M\,2-46. The spectrum available in the HASH database displays strong H$\alpha$, [N~{\sc ii}]$\lambda\lambda$6548,6583 nebular emission lines along with weak [O~{\sc iii}]$\lambda\lambda$4959,5007 lines, making the central star of M\,2-46 indiscernible in the $r$ band but visible in the $g$ band (see the PanSTARRS images in Figure\,\ref{fig: M 2-46}).  Consequently, recovering the period is compromised by the contamination of strong nebular emission in the $r$ band.  The $g$-band light curve clearly shows two peaks and troughs per orbit, with notable deviations in the depths of the two minima.  We tentatively attribute this variation to ellipsoidal modulation, with contributions from eclipses.  Follow-up observations are needed to better delineate the shape of the light curve of the central star of this quadrupolar PN. 

\emph{PHR\,J1847+0132}:  The light curve of this CSPN shows clear ellipsoidal modulation with two peaks and two troughs per orbit.  Similar to PTB\,26 and M\,2-46, such systems exhibiting significantly different minimum depths are not well understood.  One of the closest example to this curve is Th\,3-12, as reported by \citet{Jacoby2021}.  The orbital periods determined using the $r$-band and $i$-band photometric data are both close to 0.1267~d.  However, the periodicity is weak in the ZTF $g$ band, probably due to faintness of the source (g$\sim$21.1~mag), which is near the limiting magnitude of ZTF. 

\emph{Pa\,13}:  We found a consistent period of 0.3988~d in the $g$ and $r$ bands.  The light curve is dominated by two eclipses with different minima, and there seems to be some ellipsoidal modulation present.

\emph{IPHASX\,J193636.3$+$123758}: While we identified a period of 2.2150~d in the $r$ band for this object, the driving mechanism is uncertain. Due to the ambiguity of the nebula and its central star (see Section\,\ref{subsec: central star identification}), we defer any classification until more information is obtained. 

\emph{NGC\,6905}:  As the second longest variable in our sample, the light curve of NGC\,6905's central star has a period of $\sim$652~d.  While only the $r$-band photometry provides sufficient data to recover the periodicity, the long-term variation is clearly visible in both the $g$  and $r$ bands.  Its central star, HD~193949, was recently identified as belonging to the [WO] subclass of [WR] stars in a multi-wavelength study \citep{Gomez-Gonzalez2022}. Among more than one hundred [WR] stars, only a handful of them have been found to exhibit periodic photometric or radial velocity variations \citep[the central stars of PM\,1-23, NGC\,5189, NGC\,246, and Abell\,30;][]{Hajduk2010, Manick2015, Jacoby2020, Aller2020}. Thus the long-period variation of its central star is quite unusual.  However, the driving mechanism behind the observed variability remains uncertain. While we are confident that HD\,193949 is indeed a variable CSPN, definitive evidence of binarity is lacking.  Alternative explanations, such as variability arising from strong stellar winds typically observed in [WR] stars, cannot be ruled out.  Follow-up spectroscopic observations of radial velocity will be crucial in resolving this issue. 

\emph{Fr\,2-16}: Its central star exhibits light curves with a consistent period of 1.3198~d in both the $g$ and $r$ bands. Since the light curves resemble sine curves with narrower maxima, we classify the variation as due to the irradiated hemisphere of a secondary companion. However, due to the small amplitude observed in both bands, follow-up observations are recommended for confirmation.

\emph{StDr\,7}: The shape and amplitude of the light curve of this central star is typical for an irradiation effect. We identified a period of 2.6927~d in the $r$ band and a period of 2.6931~d in the $i$ band. The period reported in Table\,\ref{table: bcspne} is based on the $r$-band data, which has a much larger number of observations than the $i$ band. As illustrated in Section\,\ref{subsec: central star identification}, the periodic variable might not be the genuine central star of the nebula.

\emph{IRAS\,19581$+$3320}:  The light curve of its central star is peculiar.  We identified a consistent period of approximately 41.4~d in both the $g$ and $r$ bands.  Although the overall light curve appears highly scattered, the phase-binned version suggests a combination of ellipsoidal variability and eclipses.  This pattern is similar to the light curve of another binary central star of a PN recently identified by \cite{Wen2024}.  However, ellipsoidal variability with such a long period and considerable variation amplitude is unexpected in bCSPNe.  This object might note be a genuine PN. Therefore, we defer classifying the variability type.

\emph{Pa\,27}:  This PN central star is among the 58 binary candidates from \cite{Chornay2021gaia}, and its binarity was recently reported by \cite{Bond2024}, who found a low-amplitude sinusoidal variation with a period of 7.36~d based on the \emph{TESS}, ASAS-SN, and ATLAS photometry.  The amplitude variation suggests that the light curve is likely caused by the rotation of a spotted companion.  However, the large amplitude difference between the ATLAS and the \emph{TESS} observations might be attributed to the dilution effect caused by the relatively large measuring apertures of \emph{TESS}.  In this work we identified a period of 7.3678~d in the ZTF $g$ band, while there are insufficient observations in the $r$ and $i$ bands.  Additionally, we found a slightly deviated period of 7.3690~d from \emph{Gaia} DR3 photometry.  We classified the result as due to the rotation effect, though the phase-folded light curve from ZTF does not align with a perfect sine curve as observed in \emph{TESS} (see Figure\,\ref{fig: LCs3}).

\emph{NGC\,6833}: The central core\footnote{Here we use the term central core -- instead of central star -- of NGC\,6833, given that the central star of this compact PN is indistinguishable from the surrounding nebula even in the \emph{HST} optical images (e.g.\ \emph{HST} WFPC2/PC, prop.~ID: 6943, PI: S.\ Casertano).} of NGC\,6833 is a variable with the longest period, $\sim$950~d (see the $r$-band periodogram in Figure\,\ref{fig: periodograms}).  While such long-period photometric variation is rare in CSPNe, we were surprised to found that the variations in the three passbands ($g$, $r$, and $i$) are unusually anti-phased.  We will further discuss this object below (see Section \ref{subsec: NGC 6833}).

\emph{Kn\,26}:  The central star of Kn\,26 has the shortest period among the ZTF sample, with a period of 0.0490~d.  Similar to the morphological class of M\,2-46, Kn\,26 is another archetypal quadrupolar PN first identified in the optical bands \citep{Guerrero2013}.  We initially identified a period of 0.0490~d in the $g$ band, and 0.0467~d in the $r$ band, suggesting an influence from the window function. Upon further examination, we found that most of the outlier data points in the $r$-band light curve corresponded to the points located away from the reference position in the pointing distribution plot (see Section\,\ref{subsec: ultimate period confirmation}). These data are likely contaminated by extended nebular emission. After excluding these outlier data using a 0\farcs3 radius criterion in both the $g$ band and $r$ band, we obtained a period of 0.049~d in both bands. Therefore, we believe that the period calculated after removing the outlier data is more accurate. As the phase-folded light curve resembles a sinusoidal curve with relatively low amplitude, and the variation amplitude in the $r$-band light curve is slightly larger than in the $g$ band,  we tentatively classify its photometric variation as due to the irradiation effect.  Nevertheless, the possibility of the variation being caused by ellipsoidal modulation cannot be completely ruled out with the photometric data currently available. 

\emph{IsWe\,2}:  The light curve of this CSPN exhibits a low-amplitude and nearly sinusoidal variation. We identified a period of 0.7912~d in the ZTF $g$ band, and 0.7913~d in the $r$-band photometry. The period reported in Table\,\ref{table: bcspne} is primarily based on the $r$ band, which has a larger number of observations. The amplitude is close to the detection limit of ZTF, and based on the existing S/N, we tentatively classify its variation as dominated by the irradiation effect. However, the true orbital period will be doubled if the modulation is actually due to ellipsoidal variability.

\emph{WPS\,54}:  Its central star shows a different type of light curve compared to other variables.  We determined a consistent period of $\sim$3.45~d in the ZTF $g$, $r$ and $i$ bands, with a final period of 3.4532~d primarily based on the $r$-band photometry, which has the largest number of observations. The periodic photometric variability of this PN was previously reported by \citet{Werner2019} and \citet{Aller2024}. We also recovered a period of 3.4534~d from the ASAS-SN data, and 3.4535~d from the \emph{Gaia} DR3 photometry. The central star, PG\,0948+534, previously identified as a DA white dwarf, was observed twice in the LAMOST spectroscopic survey, on the nights of 2016 April 12 and 2022 January 03. Both of the two low-resolution ($R\sim$1800) spectra have S/N$>$50 in the $g$ band, allowing relatively reliable determination of radial velocities. The two radial velocities measured by the LAMOST pipeline differ by 45.7\,km\,s$^{-1}$, suggesting that it is a possible binary system. We tentatively classified its variation as due to the rotation effect. However, this periodicity may not necessarily be related to binarity, as \citet{Werner2019} proposed that the $\sim$3.45~d may be the rotation period of the white dwarf itself.

\emph{Kn\,51}:  We identified a period of 0.2688~d using both the $g$- and $r$-band photometric data.  The $r$-band light curve shows a nearly sinusoidal variation with a higher S/N, and its amplitude is slightly larger than that in the $g$ band. These lead us to tentatively designate its variation as due to the irradiation effect.  Additionally, we also determined a period of 0.2688~d using the \emph{Gaia} DR3 G-band photometry. This object has also been reported as a candidate variable \citep{Chornay2021gaia}. 

\begin{table*}[!ht]
\setlength{\tabcolsep}{5.pt}
\caption{\label{table: anomalous signal} ZTF Detection of the PN Central Stars with Anomalous Light Curves}
\centering
    \begin{tabular}{llcccc}
    \hline\hline
     Name & PNG & RA & Dec & HASH & Comments \\
     ~ & ~ & (J2000) & (J2000) & status\\
     \hline
        SB\,17 & 011.1-07.9 & 18:40:19.9 & -22:54:29.30 & L & periodic variable?\\
        Th\,4-11 & 011.3+02.8 & 18:00:08.8 & -17:40:43.32 & P & semi-regular variation\\
        Tan\,2 & 014.0-02.5 & 18:25:17.1 & -17:56:28.40 & T & a dip in brightness\\
        MPA\,J1852-1128 & 022.7-05.4 & 18:52:22.5 & -11:28:28.00 & L & prolonged continuous brightening\\
        K\,3-20 & 032.5-03.2 & 19:02:10.2 & -1:48:45.32 & T & long-term variation\\
        K\,3-5 & 034.3+06.2 & 18:31:45.8 & 04:05:09.13 & T & prolonged continuous brightening\\
        CTSS\,2 & 044.1+05.8 & 18:50:46.8 & 12:37:30.22 & T & prolonged continuous brightening\\
        IPHAS\,J192717.94+081429.4 & 044.3-04.1 & 19:27:17.9 & 08:14:29.40 & P & semi-regular variation\\
        StDr\,14 & 047.3+01.2 & 19:13:45.3 & 13:22:09.99 & L & a dip in brightness\\
        Hen\,1-5 & 060.3-07.3 & 20:11:56.1 & 20:20:04.42 & T & post-AGB, stellar wind?\\
        Kn\,27 & 079.8+08.6 & 19:56:00.9 & 45:23:17.02 & P & periodic with eclipsing?\\
        IRAS\,23545+6441 & 117.2+02.6 & 23:57:00.2 & 64:57:25.09 & P & semi-regular variation\\
        WeSb\,1 & 124.3-07.7 & 01:00:54.1 & 55:04:00.00 & P & eclipsing?\\
        IPHAS\,J052015.87+314938.8 & 174.5-03.0 & 05:20:15.9 & 31:49:38.78 & P & large scatter\\
        \hline
    \end{tabular}
\end{table*}

\emph{CoMaC\,2}:  We determined a clear periodic variation of the central star of this PN at 4.5877~d using the ZTF $r$-band photometric data. Doubling this period to $\sim$\,9~d yields a value that is too long to be attributed to tidal distortion effects. Since its light curve is consistent with a sinusoidal curve, we classify the variation as due to the irradiation effect.  In the LAMOST low-resolution spectrum (despite the limited S/N), we identified possible broad emission lines, which may originate from a [WR]-type central star. Given that there are only a handful of [WR]-type CSPNe exhibiting evidence of binarity, we will carry out follow-up observations of this CSPN, both high-dispersion spectroscopy and more accurate photometry. 

\emph{Kn\,133}: A consistent period of 2.4182~d was identified in both the $g$ and $r$ bands. The light curves are nearly sinusoidal with a large amplitude. Therefore, we classify this as due to presence of a cool companion, which is irradiated by the hot primary. This object has been reported as a candidate variable \citep{Chornay2021gaia}.

\emph{HaWe\,8}:  Like that of Kn\,133, the shape of the light curve of its central star is clearly suggestive of the irradiation effect, and its central star was also one of the candidate variables reported by \citet{Chornay2021gaia}. We found a consistent period of 0.9868~d in both the $g$ and $r$ bands, and 0.9867~d in the $i$ band.

\emph{WeDe\,1}: The periodic variability of this CSPN was first reported by \citet{Reindl2021}, who identified a period of approximately 0.5708~d based on $r$-band data from ZTF DR4, attributing the observed modulation to irradiation. Armed with more data, we found a consistent period of 0.5707~d in both the $g$ and $r$ band. However, the maximum brightness does not occur at phase 0.5, which deviates from expectations for purely irradiation-driven variability. Based on the shape of the light curve, we tentatively attributed the observed variations to rotational modulation. It is noteworthy that \citet{DeMarco2013} reported an $I$-band excess for this object, potentially indicating the presence of a cool companion. Additional observations would be valuable for further revealing the true nature of the central system.

\emph{PM\,1-23}:  The central star of this PN is the first to be reported as a [WR]-type central star in a close binary system.  \cite{Hajduk2010} found a period of approximately 0.63~d and tentatively attributed its photometric variation to the irradiation effect.  A follow-up study of the radial velocity \citep{Manick2015} confirmed that the orbital period is double that value.  In this work, we determined a strong periodicity of 1.2628~d based on the ZTF $g$ and $r$-band photometry.  The phase-folded light curves exhibit strong ellipsoidal variation with two peaks and two troughs per orbit. Additionally, we found that the two neighbouring peaks exhibit slight asymmetry, a feature was also present in the ASAS-SN data, which might be due to the O'Connell effect \citep{OConnell1951,Milone1968}.  An in-depth study of this object will be separately reported. 

\emph{MPA\,J0705-1224}:  We determined an orbital period of 0.1844~d from both the ZTF $g$- and $r$-band photometric data.  Similar to PHR\,J1847$+$0132, the light curves of its central star exhibit two peaks per period with markedly different depths in the two neighbouring minima, strongly suggesting an ellipsoidal modulation. 

A recent, independent study of \citet[hereafter B24]{Bhattacharjee2024} was submitted and can be used as comparison.  They reported the discovery of nine new bCSPNe using primarily ZTF forced photometry \citep{Masci2023}, whereas our work relied solely on the archival photometry (or standard photometry). ZTF Forced photometry is generally more reliable for extended sources where automatic PSF fitting may fail. However, our results indicated that standard photometry can be effective in identifying bCSPNe in most cases. For nebulae with extended but optically thin shells, the CSPN is often directly resolvable in images, and the fitting process remains largely unaffected. Conversely, when the CSPN becomes less evident or even indistinguishable due to bright nebular contamination, the process of automatic PSF source identification is significantly compromised. In such cases, the primary limitation for detecting bCSPNe is the nebular contamination itself, which obscures most observational signatures of binarity. For extremely compact PNe (e.g. NGC\,6833), the PSF fitting process performs as effectively as it does for point sources.

In comparison with the results of B24, our catalog is more comprehensive. While they applied variability metrics to pre-select candidate variables, this approach may potentially exclude some bCSPNe. In contrast, we applied our search procedure to the entire HASH database and also incorporated the ZTF $i$-band data, ensuring a more exhaustive search. Additionally, while ZTF forced photometry ensures a higher data quality in slightly extended sources, data loss remains and becomes a limitation.  Nevertheless, for the objects common in both studies, the derived orbital periods are highly consistent in values (to two decimal places), despite using the ZTF data processed with different methodologies. Among the nine new sources reported by B24, seven are included in our catalog, with H\,2-22 and M\,2-50 being the exceptions. These two objects will be detailed in the text below (see Section \ref{subsec: comparison with other studies}).

\begin{figure}[t]
\centering
\includegraphics[width=1.\columnwidth]{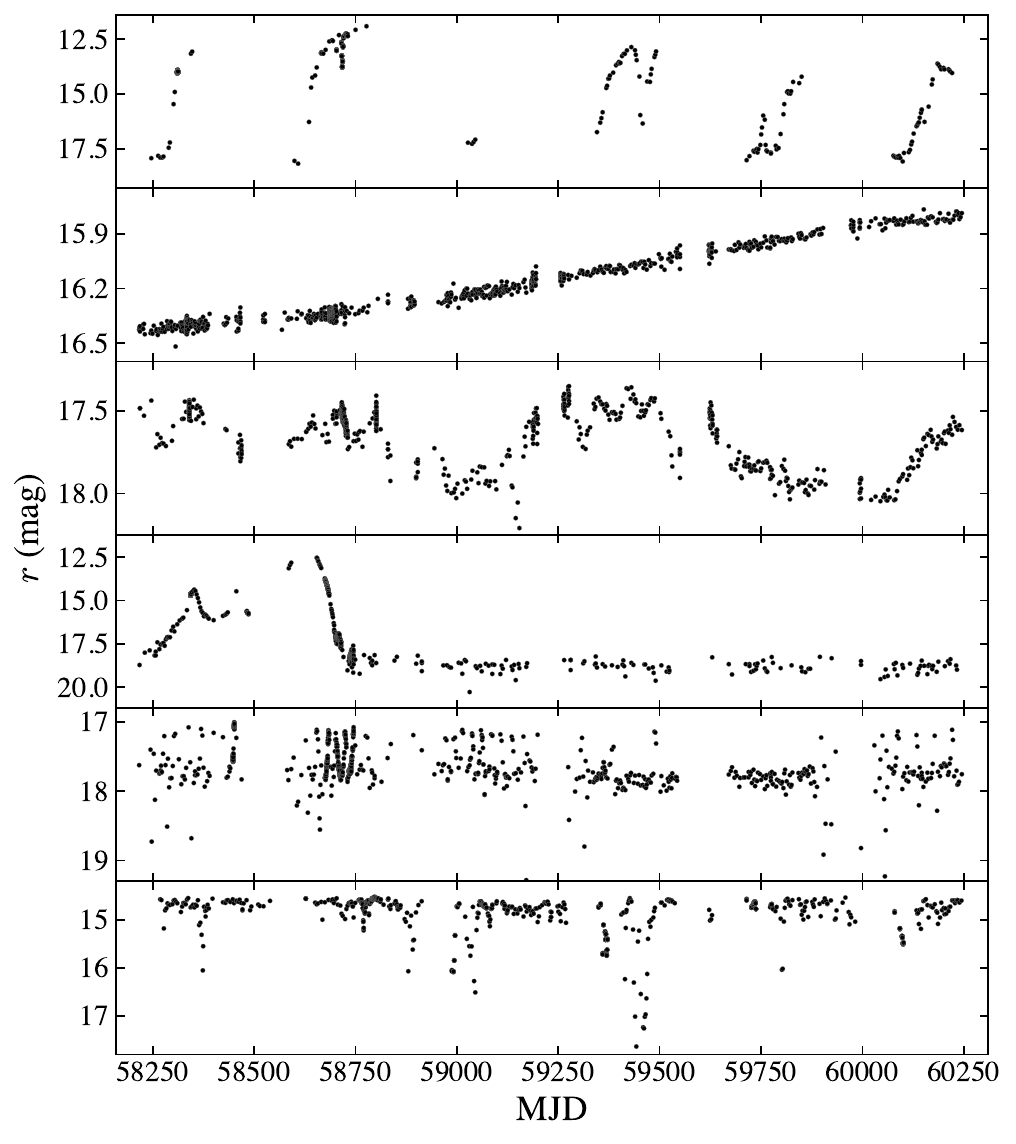}
\caption{A selection of the ZTF $r$-band light curves of the PNe, whose central stars exhibit anomalous photometric variations.  From top to bottom: SB\,17, CTSS\,2, IPHAS\,J192717.94$+$081429.4, Hen\,1-5, Kn\,27, and WeSb\,1.
 \label{fig: anomalous signal}}
\end{figure}

\subsection{Anomalous Signals} \label{subsec: anomalous signals}

Apart from the periodic variables, we also found a few central stars of PNe exhibiting anomalous signals. The light curves of these central stars are represented by large-amplitude variations with semi-regular or irregular modes, sometimes accompanied by transit-like changes in brightness. We excluded the extended PNe whose central stars are completely indiscernible in the ZTF deep reference images.  As a result, 14 CSPNe with anomalous light curves remained, which are listed in Table\,\ref{table: anomalous signal} along with basic information.  We also provide brief comments on the light curve shapes or tentative conjectures regarding the nature of these central stars.  Six representative example $r$-band light curves are depicted in Figure \ref{fig: anomalous signal}.

\begin{figure*}[t]
\centering
\includegraphics[width=2.1\columnwidth]{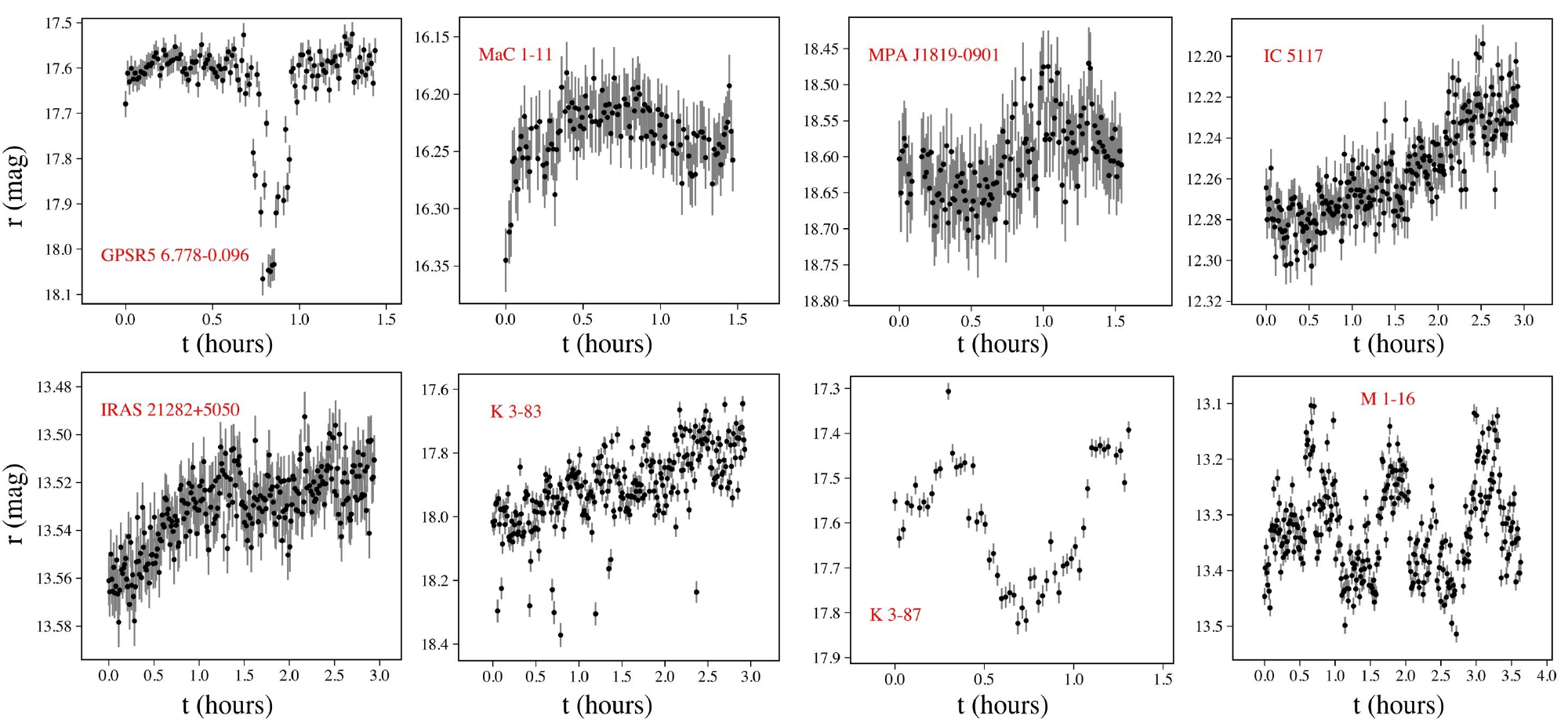}
\caption{ZTF densely-sampled light curves of the eight variables. The starting time (0.0~hr) is set to correspond with the first data point.
 \label{fig: high cadence}}
\end{figure*}

Several factors can lead to these anomalies, including accretion and mass-loss events occurring within the central system, as well as the obscuring effects of surrounding materials. For example, a detailed analysis of one of our cataloged object, WeSb\,1, can be found in \citet{Bhattacharjee2024}. Some of these phenomena could indicate the presence of a secondary star. Another plausible explanation is the contamination from PN mimics (e.g. symbiotic stars, post-AGB and other emission-line stars).  Symbiotic stars (SySt) are interacting binaries typically consisting of a white dwarf with a secondary red giant (S-type) or Mira (D-type). Their extended morphology and PN-like optical emission spectra make them the primary mimics when identifying both Galactic and extragalactic PNe. We expect that a few SySts exist among these 14 objects, as well as in the whole HASH sample.  \citet{Miszalski2009} classified the central star of the Galactic PN M\,1-34 as a likely binary based on its aperiodic variations, which are quite similar to our findings. However, we refrain from further classifying these objects in this work. Future deep spectroscopy will be necessary to first confirm their true nature as PNe.

\subsection{Other Variables from High-cadence Data}
\label{subsec: other variables from high-cadence data}

ZTF's irregular sampling pattern, combined with its  long-baseline observations, enabled the identification of several new bCSPNe exhibiting short photometric periods (e.g. Kn\,26, CGMW\,4-1723, and PHR\,J1847+0132). However, the day-cadence observations are not well-suited to resolving pseudo-periodic or aperiodic variabilities with timescales of a few hours or shorter. Fortunately, some of the CSs in our sample are located within the targeted field of the ZTF high-cadence Galactic plane survey \citep{Kupfer2021}. The continuous high-cadence observations provide higher temporal resolution and reduce the statistical errors caused by varying observing conditions, offering an opportunity to capture these fast variations. For each object there may be several segments of high-cadence observations. As mentioned in Section\,\ref{subsec: visual examination}, we have examined these densely-sampled data separately. Again, obviously extended sources were not included.

\begin{table}
\setlength{\tabcolsep}{1. pt}
\caption{\label{table: high cadence} Additional Variables Selected from High-cadence Data}
\begin{center}
    \begin{tabular}{llccc}
    \hline\hline
     Name & PNG & RA & Dec & HASH \\
     ~ & ~ & (J2000) & (J2000) & status\\
     \hline
        GPSR5\,6.778-0.096 & 006.7-00.0 & 18:01:19.0 & -23:08:43.08 & P\\
        MaC\,1-11         & 008.6-02.6 & 18:14:50.9 & -22:43:55.42 & T\\
        MPA\,J1819-0901 & 021.2+02.9 & 18:19:06.7 & -9:01:50.27 & T\\
        IC\,5117$^a$ & 089.8-05.1 & 21:32:31.0 & 44:35:47.58 & T\\
        IRAS\,21282+5050 & 093.9-00.1 & 21:29:58.4 & 51:03:59.54 & T\\
        K\,3-83 & 094.5-00.8 & 21:35:43.9 & 50:54:16.96 & T\\
        K\,3-87 & 107.4-02.6 & 22:55:07.0 & 56:42:31.14 & T\\
        M\,1-16$^a$ & 226.7+05.6 & 07:37:19.0 & -9:38:49.67 & T\\
        \hline
    \end{tabular}
    \end{center}
    \tablenotetext{a}{Slightly extended.}
\end{table}

The majority of the densely-sampled light curves exhibit a nearly ``flat'' or irregular morphology, reflecting stable brightness with minor intrinsic dispersion.  In addition to the variables included in Tables\,\ref{table: bcspne} and \ref{table: anomalous signal}, we identified eight new objects exhibiting photometric variations with regular features.  These densely-sampled light curves are presented in Figure\,\ref{fig: high cadence}.While we could not determine reliable periodicities using the complete light curves, we tentatively classified them as ``other variables" to distinguish them from true bCSPNe. The observed short-timescale photometric variations may result from stellar pulsations or fast winds associated with the central hot pre-white dwarfs. Follow-up photometric monitoring is needed to confirm their nature, and the basic information of these objects is presented in Table\,\ref{table: high cadence}.

As illustrated in Figure\,\ref{fig: high cadence}, we found direct evidence of eclipsing for GPSR5\,6.778-0.096.  However, the angular diameter of this nebula is still unclear, and the variable is located about 4\farcs1 away from the centre coordinate provided by HASH.  This ``Possible'' PN is not included in either CW21 or GS21.  Therefore, this variable source is likely a field eclipsing binary that happens to lie within the coverage of the nebula. 

We particularly highlight the case of M\,1-16, a multipolar PN with unusual, distant high-velocity jets \citep{Guerrero2020,Gomez-Munoz2023}.  We found that the high-cadence data (acquired during 2019 January) of M\,1-16 exhibits a strong periodicity at about 0.052~d, though the phase-folded light curve appears slightly scattered. This segment of high-cadence data is about 0.15~mag brighter than the regular-sampled data, and the periodicity disappeared in the second segment of high-cadence data (acquired during 2020 January).  We were unable to recover the same period when utilizing all photometric data. Therefore, we do not classify the central star as a genuine binary until more information is obtained. The observed variation may be instrumental, possibly due to the slightly extended morphology of M\,1-16. Follow-up continuous, high-cadence observations with appropriate filters are required to confirm its nature.

\begin{figure}[t]
\centering
\includegraphics[width=0.98\columnwidth]{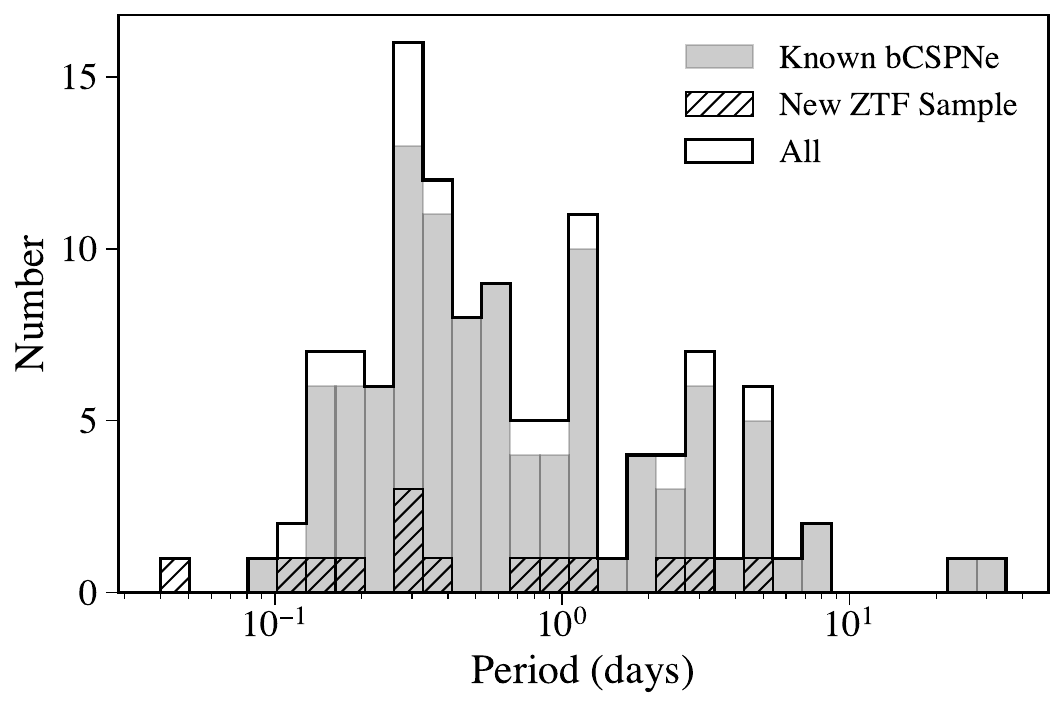}
\caption{The orbital period distribution of previously known bCSPNe (shaded), new bCSPNe discovered from ZTF (hatched).  åThe total sample is illustrated by the step plot with a thick line. 
 \label{fig: period distribution}}
\end{figure}

\section{Discussion} 
\label{sec: discussion}

\subsection{The Period Distribution} 
\label{subsec: period distribution}

We have compared the period distribution of bCSPNe from ZTF sample with those from other ground-based and space-based surveys. To ensure consistency, we only selected objects categorized into class A or B and with known orbital periods. We excluded WeDe\,1 because its periodicity likely corresponds to the rotation period. Additionally, IPHASX J193636.3+123758 was excluded due to its ambiguity of central star identification and period determination. The resultant ZTF sample comprises 29 periodic variables, with 14 being new discoveries. We also collected the 99 bCSPNe identified through photometric variations and  with known orbital periods from the online catalog contributed by David Jones. Five of the six new discoveries from \cite{Aller2024}\footnote{NGC\,1501, IC\,2149, IC\,4593, NGC\,7094, Hen\,3-1863 and WPS\,54.}, where the variability is confidently attributed to their corresponding PNe, were also considered. WPS 54 was not included because we retain a rotation period of 3.4532~d based on ZTF data, rather than half that period as indicated by \textit{TESS}. The period distribution of these bCSPNe is depicted in Figure\,\ref{fig: period distribution}. 

The known bCSPNe are clustered between 0.1 and a few days, and our newly discovered ZTF samples are similarly distributed. Although the appearance of the figure is not sensitive to the bin size, several factors could cause the distribution to deviate from the true picture. For instance, the possible confusion between irradiated and ellipsoidal variables or the influence of the aliasing effect can result in inaccurate orbital period determinations. Additionally, stellar activities like fast winds and pulsations could create mimics across the sample, further complicating the interpretation.
The overall distribution can be phenomenologically explained by a CE scenario, especially when compared with other CE survivors such as white-dwarf-main-sequence (WDMS) binaries \citep{Miszalski2009, Boffin2019}. Given  the fact that post-CE populations are strongly affected by selection effects, we refrain from making conclusive comparisons until a more unbiased sample is achieved or mature theoretical models become available.

\begin{table*}
\setlength{\tabcolsep}{4.pt}
\caption{\label{table: failed bCSPNe} Binary Central Stars of PNe with ZTF Observations but without Successfully Recovered Periodicity}
\begin{center}
    \begin{tabular}{llcccccccc}
    \hline\hline
     Name & PNG & RA & Dec & HASH & $N_{\max}$ & Period & Amplitude$^a$ & Survey & Ref$^b$ \\
     ~ & ~ & (J2000) & (J2000) & status & & (d) & (\%) & & \\
     \hline
MPA\,J1717-2356  & 000.7+08.0 & 17:17:09.0 & -23:56:29.26 & T & 534 &0.2073 & 0.7 & \textit{Kepler}/\textit{K2} & [1] \\
PHR\,J1725-2338  & 002.0+06.6 & 17:25:41.8 & -23:38:31.92 & T & 773 &0.2099 & 5.4 & \textit{Kepler}/\textit{K2} & [1] \\
PHR\,J1724-2302  & 002.4+07.0 & 17:24:53.2 & -23:02:34.04 & T & 747 &4.55  & 0.4 & \textit{Kepler}/\textit{K2} & [1] \\
Terz\,N\,1567     & 002.8+01.8 & 17:45:28.3 & -25:38:12.48 & T & 91 &0.1714 & 0.2 & \textit{Kepler}/\textit{K2} & [1] \\
PHR\,J1734-2000  & 006.2+06.9 & 17:34:13.7 & -20:00:50.54 & T & 297 &0.3367 & 1.2 & \textit{Kepler}/\textit{K2} & [1] \\
CGMW\,4-3783     & 012.1-11.2 & 18:55:04.9 & -23:28:12.70 & T & 1002 &0.6982 & 2.0 & \textit{Kepler}/\textit{K2} & [1] \\
Y-C\,2-32        & 013.7-10.6 & 18:55:30.7 & -21:49:39.65 & T & 466 &30     & 0.1 & \textit{Kepler}/\textit{K2} & [1] \\
NGC\,7293        & 036.1-57.1 & 22:29:38.5 & -20:50:13.60 & T & 181 &2.7748 & 0.17& \emph{TESS} & [2] \\ 
NGC\,7094        & 066.7-28.2 & 21:36:53.0 & 12:47:19.32 & T & 804 &4.263 & 0.25& \emph{TESS} & [3] \\ 
AMU\,1           & 075.9+11.6 & 19:31:08.9 & 43:24:58.00 & T  & 66  &2.928  & $<$0.1& \textit{Kepler}/\textit{K2} & [4]\\
Pa\,5            & 076.3+14.1 & 19:19:30.5 & 44:45:43.20 & T  & 1344 &1.12   & $<$0.1& \textit{Kepler}/\textit{K2} & [4]\\
NGC\,1501        & 144.1+06.1 & 04:06:59.4 & 60:55:14.20  & T  & 639  &3.2830& 0.20& \emph{TESS} & [3]\\
Bode\,1          & 150.9-10.1 & 03:31:12.0 & 43:54:15.48 & P  & 451  &2      & ~   &                             & [5]\\
Abell\,30        & 208.5+33.2 & 08:46:53.5 & 17:52:45.48 & T  & 767  &1.060  & 1.7 & \textit{Kepler}/\textit{K2} & [1] \\
Abell\,7$^c$     & 215.5-30.8 & 05:03:07.5 & -15:36:22.82 & T & 608  &3.0172 & 0.36& \textit{TESS} & [2] \\
                &            &            &              &   & &2.6085 & 0.31& \textit{TESS} & [2] \\ 
RWT\,152         & 219.2+07.5 & 07:29:58.6 & -02:06:37.51 & L & 557  &1.6682 & 0.16& \textit{TESS} & [2] \\ 
Me\,2-1          & 342.1+27.5 & 15:22:19.3 & -23:37:31.33 & T & 123  &22     & 0.1 & \textit{Kepler}/\textit{K2} & [1] \\ 
M\,3-39          & 358.5+05.4 & 17:21:11.5 & -27:11:37.03 & T & 137  &4.75   & 0.2 & \textit{Kepler}/\textit{K2} & [1] \\ 
Th\,3-15         & 358.8+04.0 & 17:27:10.7 & -27:43:57.60 & T & 317  &0.1508 & 3.1 & \textit{Kepler}/\textit{K2} & [1] \\
    \hline
    \end{tabular}
    \end{center}
    \tablenotetext{a}{For variables with multiple amplitudes, the largest one was adopted as the reference.}
    \tablenotetext{b}{References: [1] \cite{Jacoby2021}; [2] \cite{Aller2020}; [3] \cite{Aller2024}; [4] \cite{DeMarco2015}; [5] \cite{Bode1987}.}
    \tablenotetext{c}{Two independent photometric periodicities.}
\end{table*}

On the left-end of Figure\,\ref{fig: period distribution}, the central star of Kn\,26 currently represents the shortest-period system among all bCSPNe identified through photometric variability, with a period of 0.049~d. Nevertheless, follow-up observations are required to further confirm this period. It is also worth noting that, on the far right of the distribution, there are only two bCSPNe with orbital periods longer than ten days, both identified by \textit{Kepler}/\textit{K2}. However, if we expand the criteria to include objects detected through radial velocity variations, a few additional bCSPNe fall within this region, extending the period distribution to several years. This highlights the limitations of searching for bCSPNe using photometric observations. As the separation between the components of the binary system increases, the amplitude of brightness variations typically decreases and even become undetectable. \citet{Jacoby2021} also note that space-based observations with higher measurement precision are more likely to discover bCSPNe with longer periods. Our results align with this conclusion, as the variability observed in a few long-period variables within the ZTF sample may not be directly associated with central binarity. Although it has become clear that long-period binaries play a role in the formation and evolution of PNe comparable to that of short-period binaries \citep{Boffin2019}, this population remains poorly understood due to the limited number of known cases.

While wide binary systems have been revealed in the heart of a sample of PNe, through high-resolution space-based observations \citep{Ciardullo1999}, the identification of common proper-motion pairs based on astrometric parameters measured by \emph{Gaia} \citep{Gonzalez-Santamaria2020, Gonzalez-Santamaria2021, Ali2023}, and the discovery of a central Barium star \citep[e.g.][]{Miszalski2012, Miszalski2013, Jones2017AA}. These methods, however, provide limited or even negligible information on orbital periods. To address this gap, long-term radial velocity monitoring offers the most promising approach, though it requires a larger investment compared to photometric surveys.

\subsection{Comparison with Other Studies}
\label{subsec: comparison with other studies}

We compared the photometric periods calculated from ZTF data with those from the literature.  High consistency in the results was found, with values typically identical up to four decimal places (when available). We considered the observation epochs for all known bCSPNe that were observed by ZTF.  The same method as introduced in Section\,\ref{subsec: ultimate period confirmation} was used to combine the light curves associated with different ZTF OIDs, and the Lomb-Scargle method was used in period identification.  However, the results remained the same and no correct period was successfully recovered. In conclusion, despite there being 39 known bCSPNe with over 50 good-quality observations in at least one of the ZTF $g$, $r$, and $i$ bands, only 20 had their periodicity successfully recovered from the ZTF photometry. 

The remaining 19 bCSPNe are listed in Table \ref{table: failed bCSPNe}, with the majority discovered through space missions. Of the 36 bCSPNe discovered by \textit{Kepler}/\textit{K2} \citep{DeMarco2015, Jacoby2021}, 14 meet the criterion of $N_{\max}\geq 50$ (where $N_{\max}$ is the maximum number of good-quality observations in the $g$, $r$, and $i$ bands). However, PTB 26 is the only one for which periodicity was successfully recovered from the ZTF photometry. Additionally, the periodicity of three bCSPNe from \textit{TESS} \citep{Aller2020} observed by ZTF all fail to be recovered in this work. Of the six newly discovered bCSPNe recently reported in \cite{Aller2024}, three were observed by ZTF, with WPS 54 being the only one for which periodicity was successfully recovered.

This inconsistency might be attributed to several factors. First and foremost is the difference in amplitude sensitivity between the telescopes. More than half of the bCSPNe in Table\,\ref{table: failed bCSPNe} have amplitudes smaller than 1\%. Although the dilution effect needs to be considered for observations with relatively large pixels, such as \textit{Kepler}/\textit{K2} and \textit{TESS}, such periodicity can be easily drowned out in ground-based observations like ZTF due to less precise measurements.  In contrast, space-based observations are not affected by weather or atmospheric conditions, allowing them to detect variability with smaller amplitudes through high-precision photometry. For the objects with relatively larger amplitudes, such as PHR\,J1725-2338 and CGMW\,4-3783, we did detect peaks corresponding to their genuine orbital periods in the periodograms. However, these peaks are often inconspicuous among the crowded aliasing structures. Additionally, for a few extended objects (e.g. M\,3-39), where archival photometry is less reliable, periodicity recovery may also fail.

These highlight the great observational biases that arise when utilizing ZTF to search for bCSPNe and in estimating the binary fraction of the central systems of PNe. Simply put, if we take the results from these known bCSPNe as an average estimate, the true number of bCSPNe within the ZTF sky coverage should be at least double the number of our discoveries.

\begin{figure*}[t]
\centering
\includegraphics[width=1.0\linewidth]{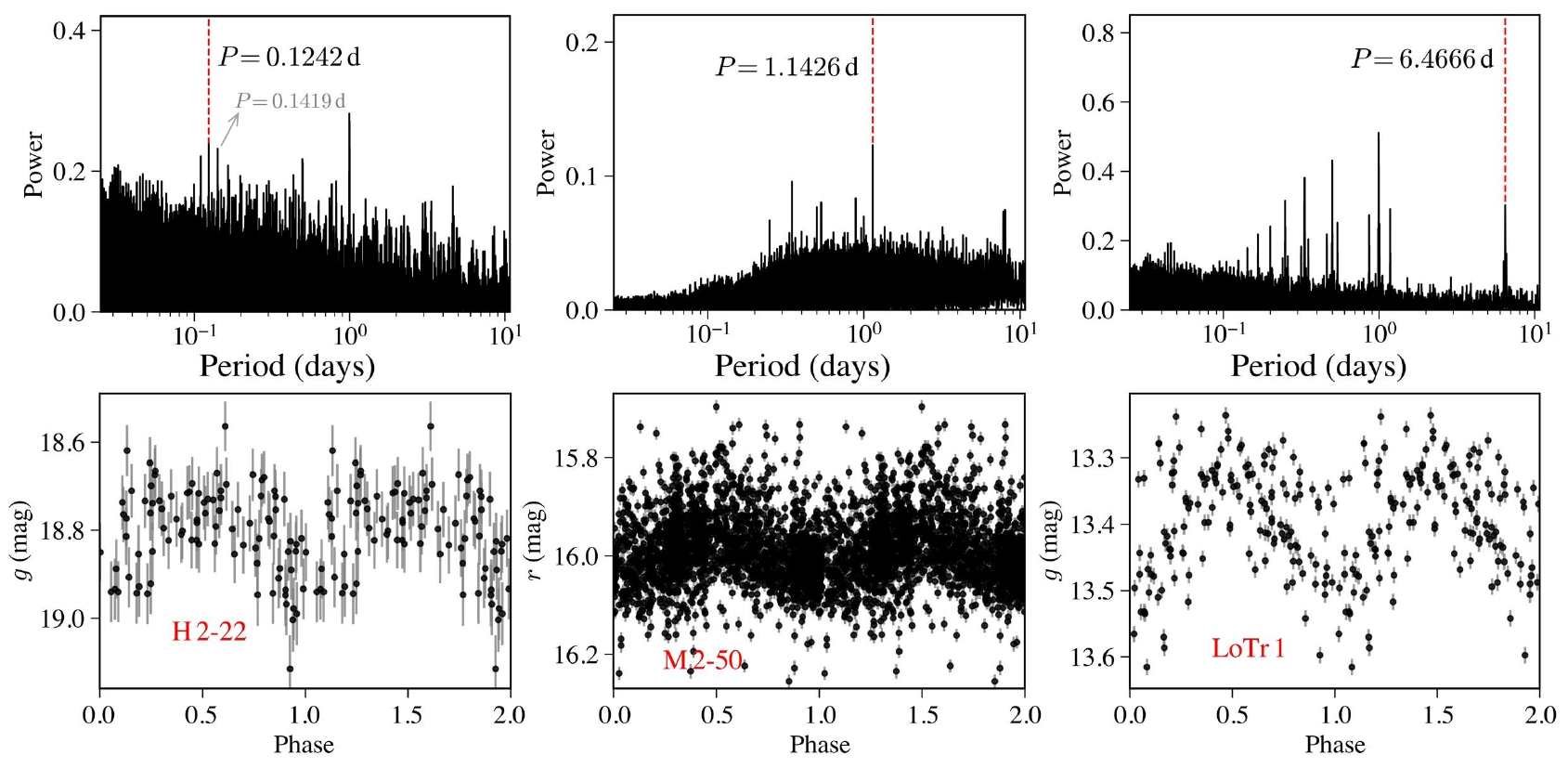} 
\caption{Periodograms (top) and phase-folded light curves (bottom) for three objects (from left to right: H\,2-22, M\,2-50, and LoTr\,1) in which periodicity was detected but ultimately excluded from classification as periodic variables. The red dotted lines in the periodograms indicate the best period after manually excluding aliasing effects. For H\,2-22, the best period reported by B24 is denoted by the gray arrow.
\label{fig: comparison}} 
\end{figure*}

Regarding the periodic variables reported in other studies but not cataloged by us, two objects from B24, H\,2-22 and M\,2-50, warrant special attention.  Figure\,\ref{fig: comparison} presents the periodograms and phase-folded light curves for H\,2-2, M\,2-50, and another representative object LoTr\,1. For H\,2-22, we detected a most probable period of 0.1242~d in the $g$ band, while B24 reported a period of 3.41~h (0.1421~d) for this PN.  For M\,2-50, we identified the same period as reported by B24.  However, the light curves for both objects appear quite scattered and noisy.  A similar issue is observed in the known bCSPNe LoTr\,1, where we derived a period of 6.4666~d in the $g$ band, consistent with the 6.5~d rotation period reported by \citet{Tyndall2013}.  These light curves were classified as periodic but with low confidence during visual inspection, leading to their periods being considered unreliable or deviating from the true values.  Objects with similar features (e.g. PM\,1-295) were excluded from our catalog and left for future investigation.

However, we note that the forced-photometry-based light curves for M\,2-50, as presented in B24, appear notably ``clean".  B24 identified the same period in the $g$ band, whereas our periodogram generated from the standard-photometry-based light curve in the same filter showed no obvious peaks. This discrepancy may be related to the slightly extended morphology of M\,2-50, whose CSPN is not clearly discernible as revealed in PanSTARRS $g$- and $r$-band images. This object exemplifies how the choice between the ZTF standard photometry and the ZTF forced photometry can yield different outcomes in identifying the bCSPNe

\begin{figure}[t]
\centering
\includegraphics[width=1.0\linewidth]{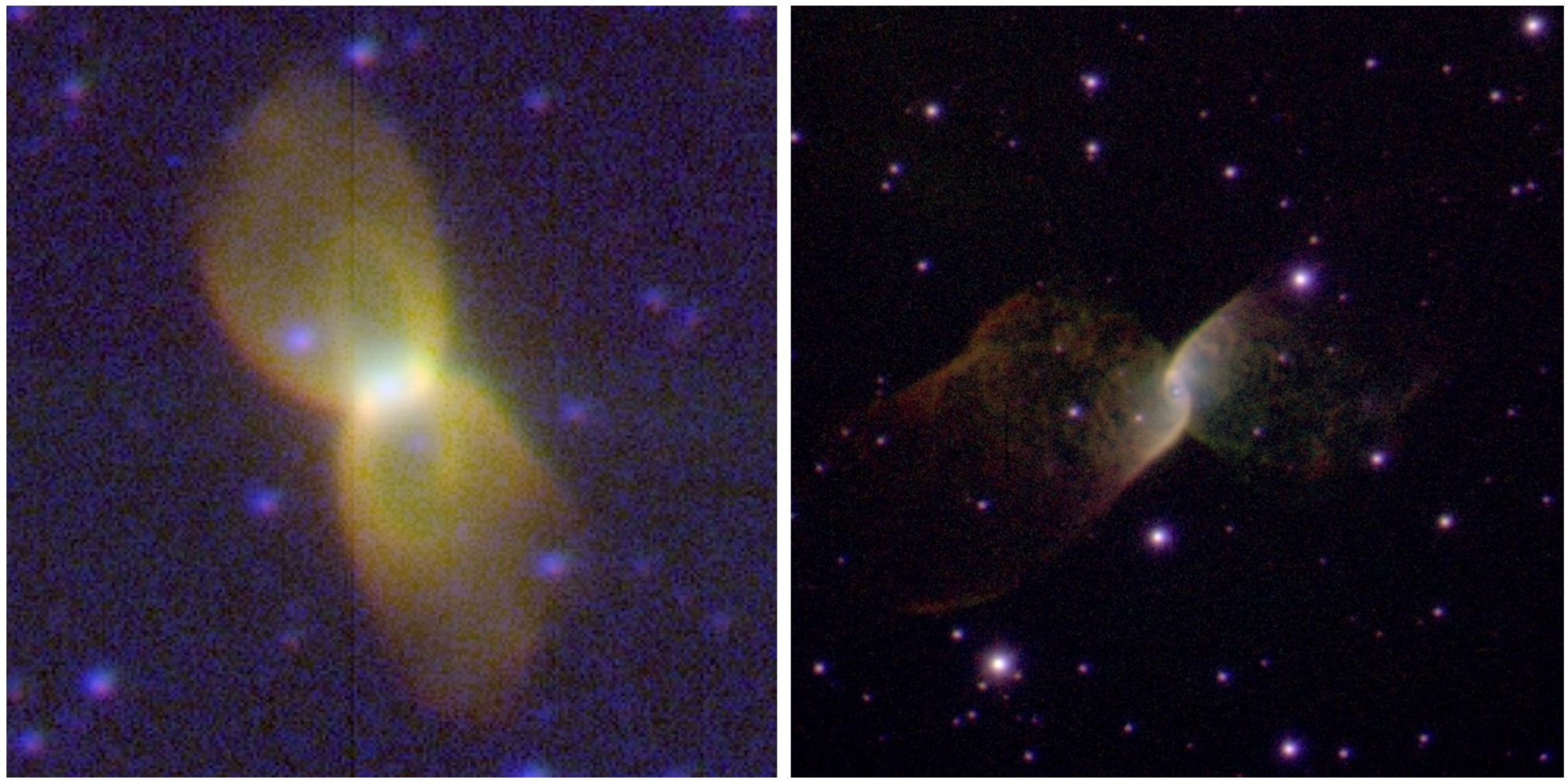} 
\caption{Color-composite images of M\,2-46 (\emph{left}, FoV$\sim$35\arcsec$\times$35\arcsec) and Kn\,26 (\emph{right}, FoV$\sim$2\arcmin$\times$2\arcmin) in the [N~{\sc ii}] (red), H$\alpha$ (green), and [O~{\sc iii}] (blue) emission lines.  North is up and east to the left.  The narrow-band optical images were obtained with the 2.56\,m Nordic Optical Telescope by \citet[][for M\,2-46]{Manchado1996} and \citet[][for Kn\,26]{Guerrero2013}. 
\label{fig: quadrupolar pne}}
\end{figure}

\subsection{Quadrupolar PNe} 
\label{subsec: quadrupolar pne}

The concept of quadrupolar PNe was first introduced by \cite{Manchado1996}. Generally, these objects feature a pronounced waist and two pairs of bipolar lobes oriented in different directions. To date, approximately ten PNe with this morphology have been detected with the aid of high-quality narrow-band imaging, though some remain controversial. A concise review can be found in \cite{Guerrero2013}. 

While binarity is believed to play an important role in the formation of quadrupolar PNe, definitive evidence has been lacking. 
\cite{Mampaso2006} reported that the `Pr\'\i ncipes de Asturias'' nebula exhibits an obvious near-infrared excess, which suggested the presence of a cool secondary.  We found that no periodic variation of the central star has been detected using ZTF photometry, though this can be attributed to several factors, such as nebular contamination or intrinsically low-amplitude variations. Another possible exception is NGC\,5189, which was classified as a quadrupolar nebula by \cite{Sabin2012}. This PN exhibit an extremely complex morphology with multiple knots and outflows \citep{Goncalves2001, Danehkar2018}.  The radial velocity and photometric variability of the central system was discovered by \cite{Manick2015} and \cite{Aller2020, Aller2024}. However, the two orbital periods are not aligned, leaving it a mystery of the true nature of the central binary system.

In the current work, we detected strong periodic photometric variations in the central stars of two quadrupolar PNe, M\,2-46 and Kn\,26.  The morphologies of the two PNe are both characterized by two pairs of well-defined bipolar lobes, as shown in the color-composite images in Figure\,\ref{fig: quadrupolar pne}.  The phase-folded light curves of M\,2-46 and Kn\,26, depicted in Figures\,\ref{fig: LCs1} (g) and \ref{fig: LCs3} (f) respectively, clearly reveal the presence of binary systems at the centres of the two representative quadrupolar PNe. 

The short orbital periods strongly suggest that the binaries at the heart of the two quadrupolar PNe have undergone a CE phase. This raises the question of what mechanism within the short-period binary system influenced the mass-loss process, resulting in such a symmetric morphology that deviates from the canonical bipolar PNe. 
The time lapse in ejection between the two pairs of lobes in Kn\,26 has been shown to be very small \citep{Guerrero2013}, a result also found in other quadrupolar PNe such as M\,1-75 \citep{Santander-Garca2010} and NGC\,6309 \citep{Rubio2015}. This implies a possible rapid change in the launching direction of the mass-loss process. Intriguingly, although M\,2-46 shares a similar morphology, the time lapse for its lobe ejection was estimated to be a few thousand years \citep{Manchado1996}. This large timescale makes the aforementioned ``rapid change'' scenario unreasonable. We have undergone follow-up spectroscopy and photometry of the two objects (Fang et al. in preparation).
Given our knowledge of the central stars remains limited, any parameters (e.g. mass ratio, orbital separation) of the central systems will provide better observational constraints.

\begin{figure}
\centering
\includegraphics[width=1.0\linewidth]{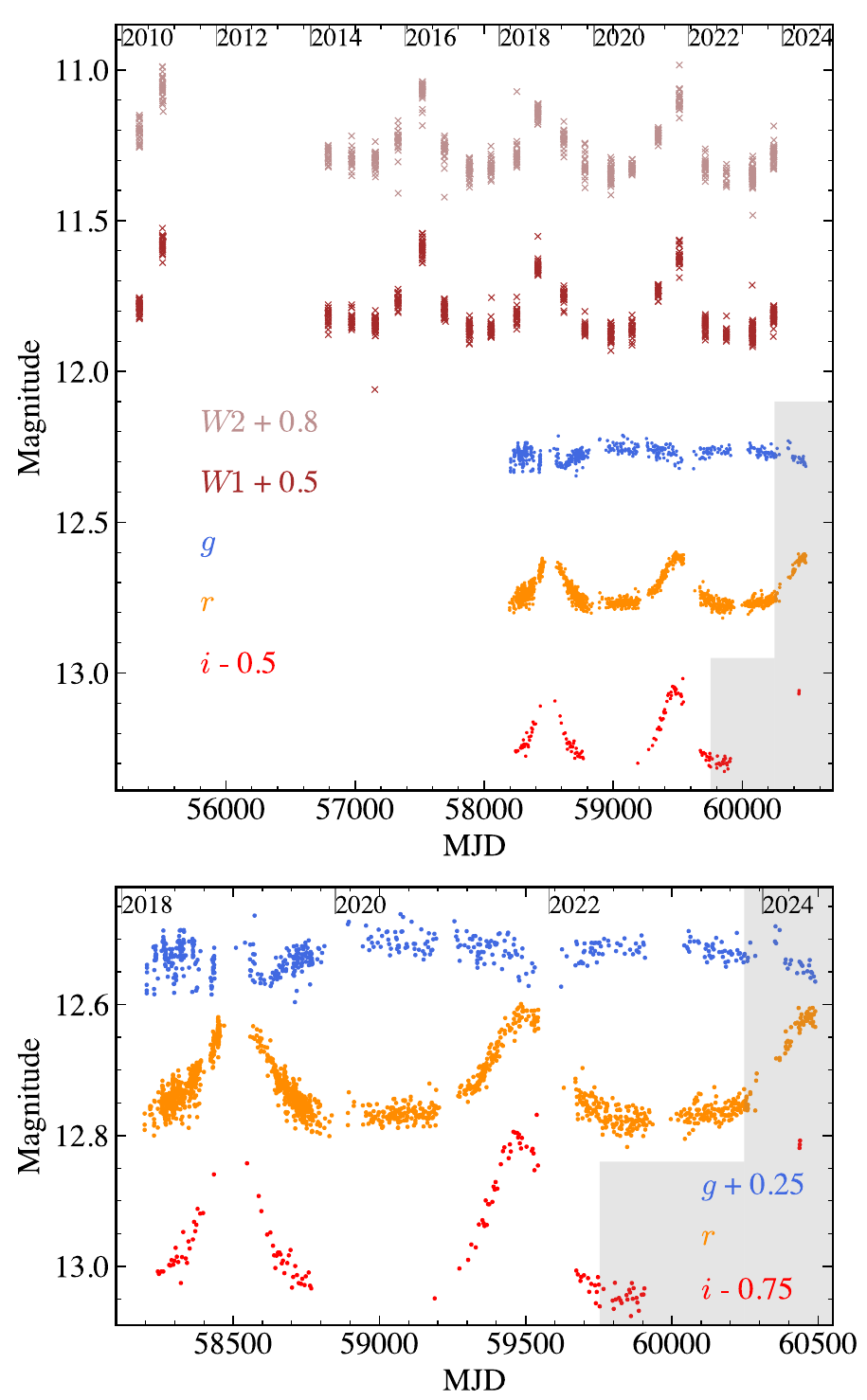} 
\caption{\emph{Top}: Light curves of NGC\,6833 in the \emph{WISE} and ZTF bands, shifted vertically for clarity in display. The gray shaded region highlights additional photometric data from the latest ZTF DR22. \emph{Bottom}: Same as the top panel but only showing the ZTF light curves in the $g$, $r$ and $i$ bands.}
  \label{fig: NGC 6833}
\end{figure}

\subsection{NGC\,6833: A Long-Period Anti-Phase Variable} 
\label{subsec: NGC 6833}

NGC\,6833 is a compact bipolar PN with jet-like structures, as revealed by the \emph{HST} optical narrow-band imaging (e.g.\ \emph{HST} WFPC2/PC, prop.~ID: 6943, PI: S.\ Casertano).  Its position, peculiar radial velocity, and low metallicity, and other spectral features all suggest that NGC\,6833 possibly belongs to an old stellar population associated with the Galactic halo, while currently experiencing the early stage of its PN evolution \citep{Wright2005, Lee2010}. 

To better illustrate the long-term variation of NGC\,6833, we collected archived infrared photometric data (W1 and W2 filters) from the \textit{Wide-field Infrared Survey Explorer} (\textit{WISE}), which comprises data releases from the \emph{ALLWISE} \citep{Wright2010,Mainzer2011} and \emph{NEOWISE-R} \citep{2014ApJ...792...30M}.  The same selection criteria as \citet{Chen2018} were adopted to exclude bad-quality photometric data.  The optical-infrared light curve of NGC\,6833 is depicted in Figure\,\ref{fig: NGC 6833}. Additionally, NGC\,6833 was undergoing its another ``outburst'' during the revision of this paper, as captured by the latest ZTF DR22. This motivates us to include the latest ZTF data, indicated by the shaded area in the figure.

Both the $r$- and $i$-band light curves exhibit repeated mysterious ``outburst'' with a period of nearly 1,000~d, lasting approximately 500~d. This long-term variation is further confirmed in the infrared band, where the W1 and W2 data capture three peaks of brightness, two of which occur almost simultaneously with those in the optical region. Intriguingly, the $g$ band shows opposite behavior compared to other passbands. Although the $g$-band data obtained around MJD 58300 is scattered, the anti-phase variation is clearly seen. While the $r$ and $i$ band reach their peak brightness, the $g$ band arrives at its valley. In addition, the amplitude of variation increases with the wavelength of filters in the optical region. These peculiar features, to the best of our knowledge, have not been previously reported to associate with PNe. Instead, such anti-phase variation with periodicity is known to be a unique characteristic of $\alpha^2$ Canum Venaticorum (ACV) variables, caused by the rotation of stars with a non-uniform surface distribution of chemical elements \citep{Faltova2021}.

Recently, \citet{Paunzen2023} reported the discovery of a variable PN, PM\,1-322, exhibiting anti-phase variation but with no periodicity. This object is not included in our sample, since it is labeled as a SySt candidate in HASH. However, it is worth noting that the one-magnitude eclipse-like event of PM\,1-322 is very similar to what we discovered in the light curves of Tan\,2 and StDr\,14 (see Section \ref{subsec: anomalous signals}). We strongly suspect that the binarity of NGC\,6833 might be the reason for such long-period anti-phase variations, though conclusive evidence is still needed. Combined with literature spectra and the GTC OSIRIS spectrum we newly obtained \citep{Fang2024}, a detailed study of this object will be presented in a future paper (Chen \& Fang, in preparation).

\section{Conclusions and Future Perspective} 
\label{sec: conclusion}

A comprehensive analysis was conducted to investigate the photometric variations of a substantial sample of the central stars of Galactic PNe, using long-term archived photometry from the ZTF.  We identified 39 PNe, whose central stars are periodic variables, including 19 new discoveries, and categorized them into three classes to assess their connection with the binarity of central systems in PNe.  The orbital or rotation periods of known bCSPNe were successfully recovered, and the majority of new periodic variables are believed to be bCSPNe, though the variability might be attributed to different driving mechanisms.  Additionally, through visual examination, we identified 14 PNe and their candidates exhibiting anomalous light curves, along with eight additional variables based on high-cadence photometric data. Notably, we found compelling evidence of binarity in two quadrupolar PNe, M\,2-46 and Kn\,26. Furthermore, NGC\,6833 was identified as a long-period anti-phase variable, exhibiting a variability pattern that had not been previously observed. 

The potential, as well as the challenges for searching bCSPNe with modern photometric surveys has been further demonstrated in this work. The ZTF sample serves as an ideal starting point to survey the strong connection between binary evolution with the formation and evolution of PNe, and the CE evolution. Additionally, follow-up radial velocity monitoring or high-precision photometric measurements for some of the more intriguing objects would be valuable in confirming their true nature.

It is believed that the continuation of existing observations, along with the future availability of revolutionary facilities such as the Large Synoptic Survey Telescope \citep[LSST;][]{Ivezi2019}, will enable us to uncover a more comprehensive and unbiased population of bCSPNe. At that point, some key statistical issues such as the binary fraction of central stars in PNe and the orbital period distribution of bCSPNe can be addressed from a more robust perspective.

\begin{acknowledgments}
We are grateful to the anonymous referee whose excellent comments and suggestions significantly improved this article. X.F.\ acknowledges support from the Youth Talent Program (2021) from the Chinese Academy of Sciences (CAS, Beijing).  This work is supported by the National Natural Science Foundation of China (NSFC) through project 11988101.  This publication is based on observations obtained with the Samuel Oschin 48-inch Telescope at the Palomar Observatory as part of the Zwicky Transient Facility (ZTF) project.  ZTF is supported by the National Science Foundation under Grants No. AST-1440341 and AST-2034437 and a collaboration including current partners Caltech, IPAC, the Oskar Klein Center at Stockholm University, the University of Maryland, University of California, Berkeley, the University of Wisconsin at Milwaukee, University of Warwick, Ruhr University, Cornell University, Northwestern University and Drexel University. Operations are conducted by COO, IPAC, and UW.
\end{acknowledgments}

\bibliography{references}{}
\bibliographystyle{aasjournal}



\end{document}